\documentclass[10pt]{article}
\usepackage[margin=0.7in]{geometry}
\usepackage{moreverb}
\usepackage{gensymb}
\usepackage[utf8]{inputenc}
\usepackage{algorithm}
\usepackage{algpseudocode}
\usepackage{amsfonts}
\usepackage{hyperref}
\usepackage{graphicx}
\usepackage[font=normalsize]{subcaption}
\usepackage{float}
\usepackage{tikz}
\usepackage{multirow}
\usepackage{nicematrix}
\usepackage{color}
\usepackage{amsmath,amsthm,bbm}
\usepackage{nicematrix}
\usepackage{xfrac}
\usepackage{makecell}
\usepackage{tabularx}
\usepackage{enumitem}
\usepackage{booktabs}
\usepackage{adjustbox}
\usepackage{caption}
\usepackage{tabularray}
\DeclareMathOperator{\Var}{Var}
\usepackage{amsmath,amsthm,amssymb, bbm}
\newtheorem{theorem}{Theorem}
\usepackage{algorithm}
\usepackage{algpseudocode}
\newtheorem{proposition}{Proposition}
\newtheorem{lemma}{Lemma}
\newtheorem{remark}{Remark}

\newcommand{\R}{R}
\newenvironment{proofsketch} {\begin{proof}[Proof sketch]} {\end{proof}}

\newcommand\BibTeX{{\rmfamily B\kern-.05em \textsc{i\kern-.025em b}\kern-.08em
T\kern-.1667em\lower.7ex\hbox{E}\kern-.125emX}}

\begin{document}

\title{A Diagnostic Tool for Functional Causal Discovery}

\author{
  Prakash, Shreya$^1$
  \and
  Xia, Fan$^2$
  \and
  Erosheva, Elena$^1$
}
\date{%
    $^1$Department of Statistics, University of Washington, Washington, USA\\%
    $^2$Department of Epidemiology and Biostatistics, University of California San Francisco, California, USA\\[2ex]%
}
\maketitle

\abstract{Causal discovery methods aim to determine the causal direction between variables using observational data. Functional causal discovery methods, such as those based on the Linear Non-Gaussian Acyclic Model (LiNGAM), rely on structural and distributional assumptions to infer the causal direction. However, approaches for assessing causal discovery methods' performance as a function of sample size or the impact of assumption violations, inevitable in real-world scenarios, are lacking. To address this need, we propose Causal Direction Detection Rate (CDDR) diagnostic that evaluates whether and to what extent the interaction between assumption violations and sample size affects the ability to identify the hypothesized causal direction. Given a bivariate dataset of size $N$ on a pair of variables, $X$ and $Y$, CDDR diagnostic is the plotted comparison of the probability of each causal discovery outcome (e.g. $X$ causes $Y$, $Y$ causes $X$, or inconclusive)
as a function of sample size less than $N$. We fully develop CDDR diagnostic in a bivariate case and demonstrate its use for two methods, LiNGAM and our new test-based causal discovery approach. We find CDDR diagnostic for the test-based approach to be more informative since it uses a richer set of causal discovery outcomes. Under certain assumptions, we prove that the probability estimates of detecting each possible causal discovery outcome are consistent and asymptotically normal. Through simulations, we study CDDR diagnostic's behavior when linearity and non-Gaussianity assumptions are violated. Additionally, we illustrate CDDR diagnostic on four real datasets, including three for which the causal direction is known.
}


\maketitle

\renewcommand\thefootnote{}
\footnotetext{Preprint. All code and materials for this paper available on \url{https://github.com/shreyap18/cddr_bivariate}. \R{} package for methods introduced in this paper available on \url{https://github.com/shreyap18/causalDiagnose}.}

\renewcommand\thefootnote{\fnsymbol{footnote}}
\setcounter{footnote}{1}

\section{Introduction}\label{sec1}

Identifying and understanding causal relations is fundamental to many scientific disciplines.
For example, in biology, it is crucial to understand the causal links in gene regulatory networks to gain insights into biological processes and diseases \cite{jiang2023signet}. While one can determine causal relationships through carefully applied interventions or randomized experiments, this is often too expensive, time-consuming, or infeasible. As a result, it is important to have reliable causal discovery methods that allow researchers to gather evidence and generate hypotheses on causal relations using observational data.

Causal discovery methods have been widely used across the sciences \cite{glymour_review_2019, zhang2012inferring, saito2023causal, kotoku2020causal, rosenstrom2012pairwise, hu2018application}. Typically, they use observational data to establish causal structure in the form of a Directed Acyclic Graph, a graph that encompasses a collection of random variables as nodes linked by directed edges. An edge connecting two variables indicates a dependency between these variables while holding all other variables constant. The assumption of acyclicity ensures that no closed loops or cycles exist within the graph. In the context of causal discovery, Directed Acyclic Graphs (DAGs) serve as models for expressing dependencies and causal links between variables. Below, we briefly review existing causal discovery methods and the extent of statistical guarantees that they provide in practice, before focusing on the simplest setting of bivariate causal relationship.

One can distinguish two main categories of causal discovery methods: constraint-based and function-based \cite{glymour_review_2019}. Constraint-based methods, such as the Fast Causal Inference (FCI) \cite{spirtes2000constructing} and the Peter Spirtes and Clark Glymour (PC) method \cite{spirtes2000constructing}, rely on the ``strong faithfulness'' assumption. This assumption states that a probabilistic graphical model captures all the conditional independence relationships present in the joint probability distribution of its variables. That is, the absence of an arrow (i.e., a directed edge) between two variables $X$ and $Y$ in the graph means that $X$ and $Y$ are conditionally independent given their parents (i.e., the variables with edges pointing into $X$ and $Y$). Constraint-based methods do not always yield a unique DAG. Instead, they usually identify a Markov equivalence class, which is essentially a group of DAGs with identical patterns of independence relations.

Constraint-based causal discovery methods do not rely on specific distributional or modeling assumptions apart from the strong faithfulness assumption. Instead, they identify causal relationships by recognizing patterns of conditional independence relations. These methods perform best with larger sample sizes that offer the statistical power needed to separate real dependencies from random noise \cite{glymour_review_2019}. Smaller datasets may lack the statistical power necessary for accurate results. Finally, constraint-based methods are not suitable for bivariate settings with no conditional independence relationships to leverage. 

On the other hand, functional causal discovery methods rely on specific functional models to identify the unique causal direction, assuming it exists. In the simplest case, for two random variables $X$ and $Y$, suppose the true causal direction is $X$ causes $Y$. Then, functional causal models are of the form:
$$Y = f(X,\epsilon, \mathbf{\theta}),$$ 
where $\epsilon$ is the noise independent of $X$, $f \in \mathcal{F}$ is the generating function, and $\mathbf{\theta}$ is the parameter set. Given $\mathbf{\theta}$, we assume that $f$ is invertible such that $\epsilon$ can be uniquely recovered from observed variables $X$ and $Y$. Functional casual discovery methods approximate the true data-generating process to allow one to identify the causal direction when certain properties---such as the independence relation between noise $\epsilon$ and the true cause $X$---are present in one direction but not in the other. This formulation can be generalized to multivariate settings.

Examples of functional causal models include Linear Non-Gaussian Acyclic Model (LiNGAM) \cite{shimizu_use_2008}, $Y = X + \epsilon$, where at most one of the noise $\epsilon$ and $X$ is Gaussian \cite{shimizu_use_2008}; Additive Noise Model (ANM), $Y = f_{AN}(X)+\epsilon$ \cite{hoyer2009nonlinear}, where $f_{AN}$ is a non-linear function; and Post-Nonlinear Causal Model (PNL), $Y = f_2(f_1(X)+\epsilon)$, where both $f_1, f_2$ are non-linear functions and $f_2$ is assumed to be invertible \cite{zhang2006extensions, zhang2012identifiability}. 
Shimizu et al. ~\cite{shimizu2006linear, shimizu2011directlingam}, Hoyer et al. ~\cite{hoyer2009nonlinear}, and Zhang et al. ~\cite{zhang2012identifiability} prove that directionality is identified and provide methods to estimate the causal direction for LiNGAM, ANM, and PNL functional causal models, respectively.

Functional causal discovery methods are widely applied in various fields, including neuroscience, epidemiology, psychology, and other domains within the medical, social sciences, and humanities, as demonstrated in real-world data applications \cite{saito2023causal, kotoku2020causal, rosenstrom2012pairwise, hu2018application}. For instance, in a study by Saito et al.~\cite{saito2023causal}, LiNGAM was employed to identify the factors contributing to nitrogen oxide generation by analyzing observational data from a coal-fired power plant. Similarly, Rosenstr\"{o}m et al.~\cite{rosenstrom2012pairwise} used LiNGAM to determine the causal relationship between sleep and depression. In another application, Hu et al.~\cite{hu2018application} used ANMs for causal discovery analysis with genetic data. Additionally, Song et al. ~\cite{song2017tell} applied ANM and PNL methods to bivariate data to fields such as geography, biology, physics, economics, and others. Few other applications of PNL method exist, likely due to ongoing work for efficient estimation.

Even though functional causal discovery methods rely on strong assumptions that are often questionable in practice, testing for assumption violations in causal discovery often comes as an afterthought. Diagnostics designed to assess both assumption violations and their impact on causal discovery outcomes do not exist. Thus, in functional causal discovery, to assess the validity of distributional assumptions, researchers commonly turn to post-hoc diagnostic tests designed for estimation rather than causal discovery. For example, for LiNGAM, a common approach involves visually inspecting scatter plots and examining the significance of a quadratic term in a regression model to check for linearity \cite{rosenstrom2012pairwise, rosenstrom2020distribution,song2017tell}. Additionally, tests such as the Lilliefors or Kolmogorov-Smirnov are employed post-hoc to evaluate non-Gaussianity \cite{rosenstrom2012pairwise, rosenstrom2020distribution, xu2014pooling}. These post-hoc diagnostic tests, however, can be misleading for several reasons. First, they operate based on LiNGAM, (e.g. $Y = X + \epsilon, \epsilon \perp X$), which assumes both linearity and non-Gaussianity. If LiNGAM model is flawed due to violations of either linearity or non-Gaussianity assumptions, the Lilliefors or Kolmogorov-Smirnov tests might not effectively detect these violations because they rely on the very model they aim to validate. Second, tests for non-Gaussianity and other distributional assumptions can be sensitive to sample size. Smaller samples might not have enough statistical power to detect violations, meaning that underpowered tests could produce non-significant results even when the null hypothesis does not hold, potentially leading to false negatives. This could lead to erroneously concluding that an assumption holds when it does not, which could result in the wrong directionality estimate. The aforementioned issues with post-hoc diagnostic tests are not specific to LiNGAM but are also applicable to other functional causal discovery methods.

Causal discovery methods can withstand a degree of bias in their point estimates as long as the overall causal direction is correctly identified. That is because, although assumption violations can introduce biases in estimates in functional causal discovery akin to those encountered in standard estimation problems (e.g., coefficient estimates in linear regression), these biases do not necessarily compromise the validity of causal direction identification. Thus, diagnostic tools for causal discovery should differ from conventional diagnostic tools for estimation because the target is the correct direction rather than unbiased estimates. Instead, diagnostic tools for causal discovery should allow researchers to evaluate assumption violations that may affect a causal discovery method's ability to identify the correct causal direction. When evaluating how assumption violations affect causal discovery, it is critical to account for the interplay between assumption violations and sample size. While standard estimation problems are primarily concerned with the risks associated with small sample sizes leading to indeterminate results, causal discovery poses additional challenges with increased sensitivity to minor assumption violations at larger sample sizes that may obscure the causal direction in some cases. Hence, to effectively apply causal discovery methods, causal discovery diagnostic tools need to be flexible enough to allow researchers to understand the relationship between sample size and assumption violations, potentially more intricate compared to classic statistical estimation settings. 

In this paper, we propose Causal Direction Detection Rate (CDDR) diagnostic for causal discovery to allow researchers to evaluate whether and to what extent the interaction between assumption violations and sample size affects the ability to identify the correct causal direction. This diagnostic can be applied to any bivariate functional causal discovery method. We demonstrate CDDR diagnostic for LiNGAM-based methods in the linear bivariate setting, focusing on its ability to detect violations of linearity and non-Gaussianity---two crucial assumptions for identifying causal direction--- assuming all other assumptions are met. We apply the diagnostic to both conventional LiNGAM and our proposed test-based approach. Compared to LiNGAM, the test-based approach directly incorporates hypothesis testing which makes it more informative when paired with CDDR diagnostic.

Our paper follows this structure: we first motivate and introduce Causal Direction Detection Rate (CDDR) diagnostic. We then introduce the new test-based approach and describe how to apply CDDR diagnostic to both LiNGAM and the test-based causal discovery approaches. Next, we demonstrate the use of CDDR diagnostic on simulations featuring various degrees of non-Gaussianity and linearity assumption violations. Finally, we present CDDR diagnostic analysis for four real bivariate datasets, three of which have known causal directions. We conclude the paper with a discussion of open questions for future research.

\section{Methods}

\begin{figure}
    \centering
    \includegraphics[width = 150mm]{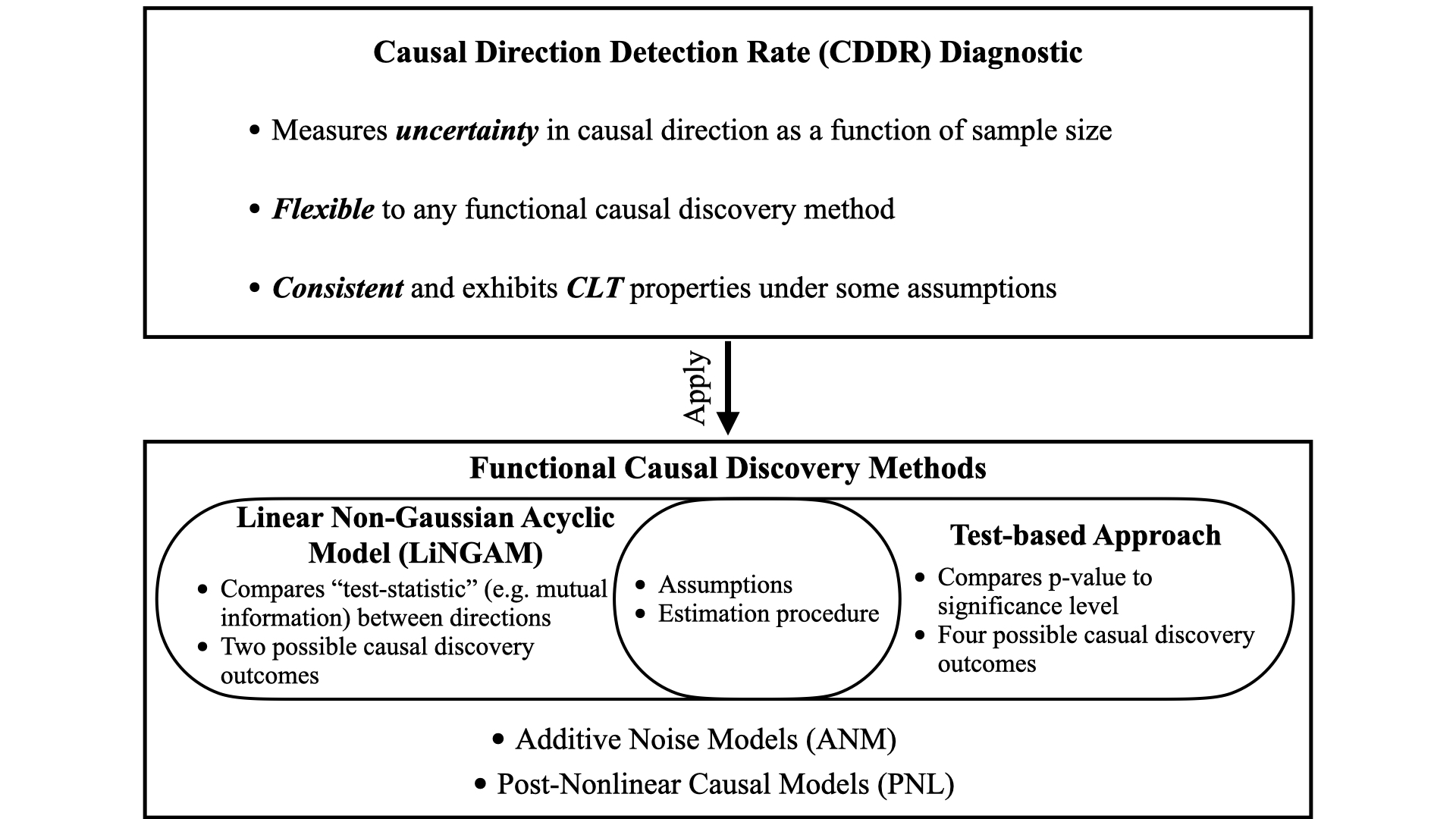}
    \caption{Schematic outline of contributions: Development of a new diagnostic for causal discovery and its illustration.}
    \label{fig:cddr_big_picture}
\end{figure}

In Section \ref{subsec:CDDR_gen}, we formally define CDDR diagnostic. In Section \ref{subsec:CDDR_bivariate}, we apply CDDR diagnostic to the bivariate setting. We summarize schematically the contributions of this paper in Figure \ref{fig:cddr_big_picture}.

\subsection{Causal Direction Detection Rate (CDDR) Diagnostic}
\label{subsec:CDDR_gen}

Given a bivariate dataset of size $N$ on a pair of variables $X$ and $Y$, where a causal direction exists, various causal discovery methods may yield different possible outcomes. For instance, LiNGAM's two possible outcomes are either $X$ causes $Y$ or $Y$ causes $X$. Another method, however, might have an inconclusive result---suggesting that neither direction is more likely than the other---as a possibility in addition to the two directionality outcomes. For a given hypothesized causal direction, which could either be based on prior knowledge or be the outcome of a causal discovery method of choice on the dataset, our diagnostic tool can be used to determine whether the hypothesized direction can be detected with confidence.

We define Causal Direction Detection Rate (CDDR) diagnostic as a plot of the probabilities/rates of detecting each possible causal discovery outcome as a function of subsample sizes less than or equal to $N$. Among the set of possible casual discovery outcome rates plotted in CDDR diagnostic, we pay specific attention to the rate of detecting $X$ causes $Y$ which is the hypothesized direction. We use the procedure in Algorithm \ref{alg:dir_rate} to estimate the rate of detecting each possible causal discovery outcome as a function of subsample size. In this procedure, we first define a vector of subsamples sizes, $(n_1,n_2,\cdots,n_j,\cdots,N)$, for $n_j$ in $j = 1, \cdots,J, J \leq N$, where $n_1<n_2<\cdots<n_j<\cdots<N$. 
For each subsample size $n_j$ in this vector, we generate $S$ subsamples of size $n_j$ by randomly sampling data points from the dataset with replacement. Next, we apply the causal discovery method of choice to each subsample. Using the casual discovery outcomes from all $S$ subsamples, we estimate the proportion of each casual discovery outcome for each subsample size $n_j$. In this procedure, we subsample with replacement to reduce the dependencies between subsamples for more reliable estimates of population parameters, aligning with bootstrap theory \cite{EfroTibs93}.

\begin{algorithm}
\caption{General Procedure to Estimate and Visualize Causal Direction Detection Rate (CDDR) Diagnostic} \label{alg:dir_rate}

\begin{enumerate}
    \item Set $N$, the sample size of the dataset
    \item Set a vector of subsamples sizes $(n_1,n_2,\cdots,n_j,\cdots,N)$, for $n_j$ in $j = 1, \cdots,J, J \leq N$, where $n_1<n_2<\cdots<n_j<\cdots<N$. For example, the vector could be $(20, 30, 40, \cdots, N)$
    \item For each subsample size $n_j$:
    \begin{enumerate}
        \item Repeat the following $S$ times:
        \begin{enumerate}
            \item Sample $n_j$ data points from dataset with replacement to form subsample $S_{n_j,s}$, where $s \in \{1, \cdots, S\}$.
            \item Run causal discovery method for subsample $S_{n_j,s}$. 
        \end{enumerate}
        \item Using all $S$ subsamples, calculate the proportion of each possible causal discovery outcome from the causal discovery method.
    \end{enumerate}
\end{enumerate}
\end{algorithm}

In Theorem \ref{thm:clt}, we prove that as the number of subsamples (S) and dataset size (N) increase to infinity, the estimated causal discovery outcome rates are consistent and asymptotically normal. This allows us to add point-wise confidence intervals to our CDDR diagnostic. In Remark \ref{rem:conditions}, we discuss the implications of using Theorem \ref{thm:clt} in practice to obtain these confidence intervals.

\begin{theorem}
\label{thm:clt}
    Let $n$ be a fixed subsample size and $s \in \{1, \cdots, S\}$ denote the subsample number. Then let $W_{n, s}$ be an indicator for the casual discovery method choosing the hypothesized direction ($X$ causes $Y$) for subsample $S_{n,s}$. Then, we denote $\hat{p}_n \equiv \frac{1}{S}\sum^S_{s=1}W_{n,s}$ as the estimated rate of detecting $X$ causes $Y$ for subsample size $n$.
    
    Let $\{k_S\}_{S=1}^\infty$ and $\{m_S\}_{S=1}^\infty$ be two sequences of natural numbers such that $k_S < m_S$ and $$\frac{k_S}{m_S} \rightarrow \gamma \in [0,1).$$
    If for a fixed subsample size $n$, $\lim_{N,S \to \infty} S n\log\left(1-\frac{(nS)(nS - S)}{N(N-nS)^{n-1}}\right) = 0$ and $N > Sn$, then as $S \to \infty$, $N \to \infty$, we have 
    \begin{align}
        \hat{p}_n &\overset{p}{\to} p_n, \text{ and} \\
        \frac{1}{\sqrt{m_S}} \sum_{s=1}^{m_S} \left(\frac{W_{n,s} - E[W_{n,s}]}{\sqrt{\sfrac{\Var[W_{n,s}]}{S}}} \right) &\overset{D}{\to} N(0, 1).
    \end{align}
    This holds for all estimated causal discovery outcome rates in Algorithm \ref{alg:dir_rate}.
\end{theorem}

\begin{proofsketch}
    Without loss of generality, let the hypothesized direction be $X \rightarrow Y$. First, we prove that we have asymptotically independent and identically distributed (i.i.d) data if our sufficient conditions hold. Then, we use this to fit the conditions given by Soloveychik in their statement of the Central limit theorem (CLT) for Symmetric Exchangeable RVs \cite{soloveychik2020central}. The more restrictive condition in practice is $N > Sn$. This condition is based on the intuition that for our subsamples to have no overlap asymptotically, the size of the data pool ($N$) must be larger than the product of the number of subsamples ($S$) and the subsample size ($n$). The product ($Sn$) represents the total data drawn from the data pool in our subsampling procedure when there is no overlap in data points. See the appendix for a full proof. 
\end{proofsketch}

\begin{remark}
    \label{rem:conditions}
    In practice, it may not be realistic for the dataset size $N$ to be larger than $Sn$ for a large $S$. This is because CDDR diagnostic considers subsample sizes up to $N$, and $N$ itself may not be large due to limited available data. 
    We use simulations to study how reliable our estimates of variability and confidence intervals are in practice, both when the sufficient conditions in Theorem \ref{thm:clt} do and do not hold. These simulations compare the estimated variability and confidence intervals of the causal discovery outcome rates with the true variability and quantiles. We find the differences to be negligible, regardless of whether the sufficient conditions in Theorem \ref{thm:clt} hold or not. See appendix for details. 
\end{remark}

The CDDR diagnostic definition and properties presented are general enough such that it can be applied to any bivariate functional causal discovery method.

\subsection{Demonstrating CDDR Diagnostic for Bivariate Causal Discovery}
\label{subsec:CDDR_bivariate}

We apply CDDR diagnostic to two methods in the linear bivariate setting: LiNGAM and our proposed test-based causal discovery approach. We introduce the test-based approach in \ref{subsubsec:test-based}. Unlike LiNGAM, which deterministically decides between two directionality outcomes based on an independence measure comparison, the test-based approach considers four causal discovery outcomes based on hypothesis testing. In Section \ref{subsubsec:CDDR_test-based}, we demonstrate how to apply CDDR diagnostic to both the test-based approach and LiNGAM. 

\subsubsection{Introducing the Test-based Approach for Linear Bivariate Causal Discovery}
\label{subsubsec:test-based}

We propose a test-based approach that extends LiNGAM by using hypothesis testing while maintaining the same assumptions. This extension improves upon LiNGAM by providing some statistical guarantees through hypothesis testing. 

Assume there exists a causal direction. In the bivariate case, both LiNGAM and test-based approaches are tasked with selecting between two linear causal models:

\begin{enumerate}
    \item $X \rightarrow Y$, represented as:
    \begin{equation}
        \label{eq:lingam_xy}
        Y = \beta X + \epsilon, \:\: X \perp \epsilon
    \end{equation}
    \item $Y \rightarrow X$, represented as:
    \begin{equation}
    \label{eq:lingam_yx}
    X = \gamma Y + \eta, \:\: Y \perp \eta
\end{equation} 
\end{enumerate}

The test-based approach translates this task into the following hypotheses of interest:

\begin{equation}
\label{eq:h_arrow}
    H = 
\begin{cases}
    H_Y^0: X \rightarrow Y, \:\: H_Y^1: Y \rightarrow X \\
    H_X^0: Y \rightarrow X, \:\: H_X^1: X \rightarrow Y
\end{cases}
\end{equation}

To determine causal direction, the test-based approach relies on the same five key assumptions as LiNGAM \footnote{Shimizu et al. demonstrate the necessity of these assumptions for causal identifiability in their work \cite{shimizu_use_2008}. They illustrate that the higher-order moment structures of Equations (\ref{eq:lingam_xy}) and (\ref{eq:lingam_yx}) are distinguishable. In the linear setting, Shimizu et al. prove that the causal direction is unidentifiable if and only if the error terms $X$ and $Y$ are Gaussian \cite{shimizu2011directlingam}. If all involved random variables are Gaussian, they construct an example using the properties of Gaussian random variables to show that both $X$ and $\epsilon$, as well as $Y$ and $\eta$, are uncorrelated and thus independent. On the other hand, they demonstrate that if both Equations (\ref{eq:lingam_xy}) and (\ref{eq:lingam_yx}) hold, they employ the Darmois-Skitovich theorem \cite{skitovic_1954, skitovic_1962, darmois_1953} to establish that $X$, $Y$, and the error terms are Gaussian.}:

\begin{enumerate}
    \item There exists a linear relationship between $X$ and $Y$,
    \item At least one of the variables $X$, $Y$, or the error term is non-Gaussian,
    \item $X$ and $Y$ have an acyclic relationship,
    \item $X$ and $Y$ are unconfounded, meaning they have no common causes
    \item All involved random variables are independent and identically distributed (i.i.d).
\end{enumerate}

Under these assumptions, we have:

$$X \rightarrow Y \Rightarrow Y = \beta X + \epsilon, $$
$$Y \rightarrow X \Rightarrow X = \gamma Y + \eta.$$
We can now reformulate $H$ from Equation (\ref{eq:h_arrow}) as follows: 

\begin{equation}
    \label{eq:h_indep}
    H^* = 
    \begin{cases}
        H_Y^0: X \perp \epsilon, H_Y^1: X, \epsilon \text{ dependent } \\
        H_X^0: Y \perp \eta, H_X^1: Y, \eta \text{ dependent }
    \end{cases}
\end{equation}
Analogous to LiNGAM, one can select which independence test to use in Equation (\ref{eq:h_indep}) for the test-based approach. We propose using the independence and goodness-of-fit test introduced by Sen and Sen \cite{sen_testing_2014}. This test, designed for linear regression, simultaneously checks for independence between the error and the predictor terms using the Hilbert-Schmidt independence criterion as well as the goodness of fit of the parametric model (Equation (\ref{eq:h0})). This dual evaluation makes the test-based approach, when paired with the independence and goodness-of-fit test introduced by Sen and Sen \cite{sen_testing_2014}, particularly sensitive to assumption violations. In practice, for the Sen and Sen test, bootstrap can be used to obtain consistent $p$-value estimates\cite{sen_testing_2014}.

\begin{equation}
\label{eq:h0}
    H_0: X \perp \epsilon, \; \;\text{relationship between } X \text{ and } Y \text{ is linear}
\end{equation}

The test-based approach improves upon LiNGAM in several ways. Instead of solely comparing values of independence measures or test statistics, the test-based approach obtains two $p$-values corresponding to hypothesis tests in each direction. These hypothesis test results convey uncertainty in the direction estimate and provide insights into assumption violations. When assumptions hold, the two $p$-values are perfectly correlated, meaning that, with enough samples, rejecting in one direction would imply a failure to reject in the other. Therefore, given no assumption violations and a sufficiently large sample size, the test-based approach would provide the same information as LiNGAM. 
However, strict assumptions such as linearity are commonly violated in practice. Under assumption violations, the two $p$-values would not be perfectly correlated, and the test-based approach would convey this information by either rejecting or failing to reject in both directions, resulting in an inconclusive directionality. In contrast, many LiNGAM implementations do not output an inconclusive directionality but instead select one of two causal directions, either $X \to Y$ or $Y \to X$, as the outcome.

\begin{table}
    \centering
    \caption{Causal discovery outcomes and their interpretation under the test-based approach, using the Sen and Sen \cite{sen_testing_2014} test for independence and goodness of fit.}
    \begin{tabular}{p{5cm}|p{10cm}}
    \hline
         \textbf{Test Outcome} & \textbf{Interpretation} \\
         \hline
         {Reject both $H_Y^0$ and $H_X^0$.} & {\textit{Inconclusive} outcome due to possible linearity assumption violations. The null $H_0$ (\ref{eq:h0}) would not hold in both directions in this case.}
         \\
         \hline
         {Fail to reject both $H_Y^0$ and $H_X^0$.} & {\textit{Inconclusive} outcome due to possible identifiability issues or small sample size. Direction is not identifiable if non-Gaussianity assumption violations are present \cite{shimizu2011directlingam}.}\\
         \hline
         {Reject $H_Y^0$ and fail to reject $H_X^0$.} & {Favors the direction $Y \rightarrow X$.}\\
         \hline
         {Fail to reject $H_Y^0$ and reject $H_X^0$.} & {Favors the direction $X \rightarrow Y$.}\\
         \hline
    \end{tabular}
    \label{Tab:sen_assume}
\end{table}

The effects of assumption violations on the test-based approach using the Sen and Sen test \cite{sen_testing_2014} are outlined in Table \ref{Tab:sen_assume}. In summary, if both null hypotheses are rejected, the outcome is inconclusive. This could be due to a linearity assumption violation, as the null (\ref{eq:h0}) would not hold in this case. On the other hand, failing to reject both null hypotheses also results in an inconclusive outcome due to small sample size or possible identifiability issues. Based on Shimizu et al.'s findings~\cite{shimizu2011directlingam}, the direction is not identifiable if the non-Gaussianity assumption is violated . 
\renewcommand{\arraystretch}{1.2}

\subsubsection{Demonstrating CDDR Diagnostic for the Test-based Approach and LiNGAM}
\label{subsubsec:CDDR_test-based}

While the test-based approach provides more statistical guarantees than LiNGAM with hypothesis testing, it still has its limitations as it does not reflect the interplay between sample size and assumption violations. As a result, both LiNGAM and the test-based causal discovery approaches will benefit from using CDDR diagnostic. CDDR diagnostic summarizes the joint distribution of the possible causal discovery outcomes as a function of sample size. Specifically, for the test-based approach, CDDR diagnostic summarizes the joint distribution of $p$-values from both hypothesis tests as a function of sample size, accounting for how sample size can affect hypothesis test outcomes in each direction differently.

We demonstrate the application of CDDR diagnostic to LiNGAM and the test-based approach. In both cases, we first use Algorithm \ref{alg:dir_rate} to estimate the rates of causal discovery outcomes for subsample sizes smaller than the observed. For LiNGAM, the procedure estimates (1) the probability of determining $X$ causes $Y$ and (2) the probability of determining $Y$ causes $X$. For the test-based approach, the procedure estimates the probabilities corresponding to detecting each outcome from the set of hypothesis tests outlined in \ref{eq:h_indep}: (1) the probability of rejecting both $X \rightarrow Y$ and $Y \rightarrow X$, (2) the probability of failing to reject both $X \rightarrow Y$ and $Y \rightarrow X$, (3) the probability of rejecting $X \rightarrow Y$ and failing to reject $Y \rightarrow X$, and (4) the probability of failing to reject $X \rightarrow Y$ and rejecting $Y \rightarrow X$. We then use CDDR diagnostic to visually compare the estimated rate of the hypothesized direction (e.g., $X \rightarrow Y$) against the other estimated outcome rates across several subsample sizes less than or equal to $N$.

\section{Simulations}
\label{sec:sims}

In this section, we study the behavior of CDDR diagnostic when applied to LiNGAM and the test-based approach under varying levels of linearity and non-Gaussianity assumption violations. We describe the simulation setup and the interpretation of CDDR diagnostic in practice in Section \ref{subsec:sims_setup}. We discuss the simulation results for varying levels of linearity and non-Gaussianity assumption violations in Sections \ref{subsec:lin_sims} and \ref{subsec:non_gauss_sims}, respectively.

\subsection{Simulation Setup}
\label{subsec:sims_setup}

\begin{table}
    \centering
    \caption{Summary of how to read and interpret CDDR diagnostic for example of LiNGAM and test-based approach}
    \includegraphics[width=\textwidth]{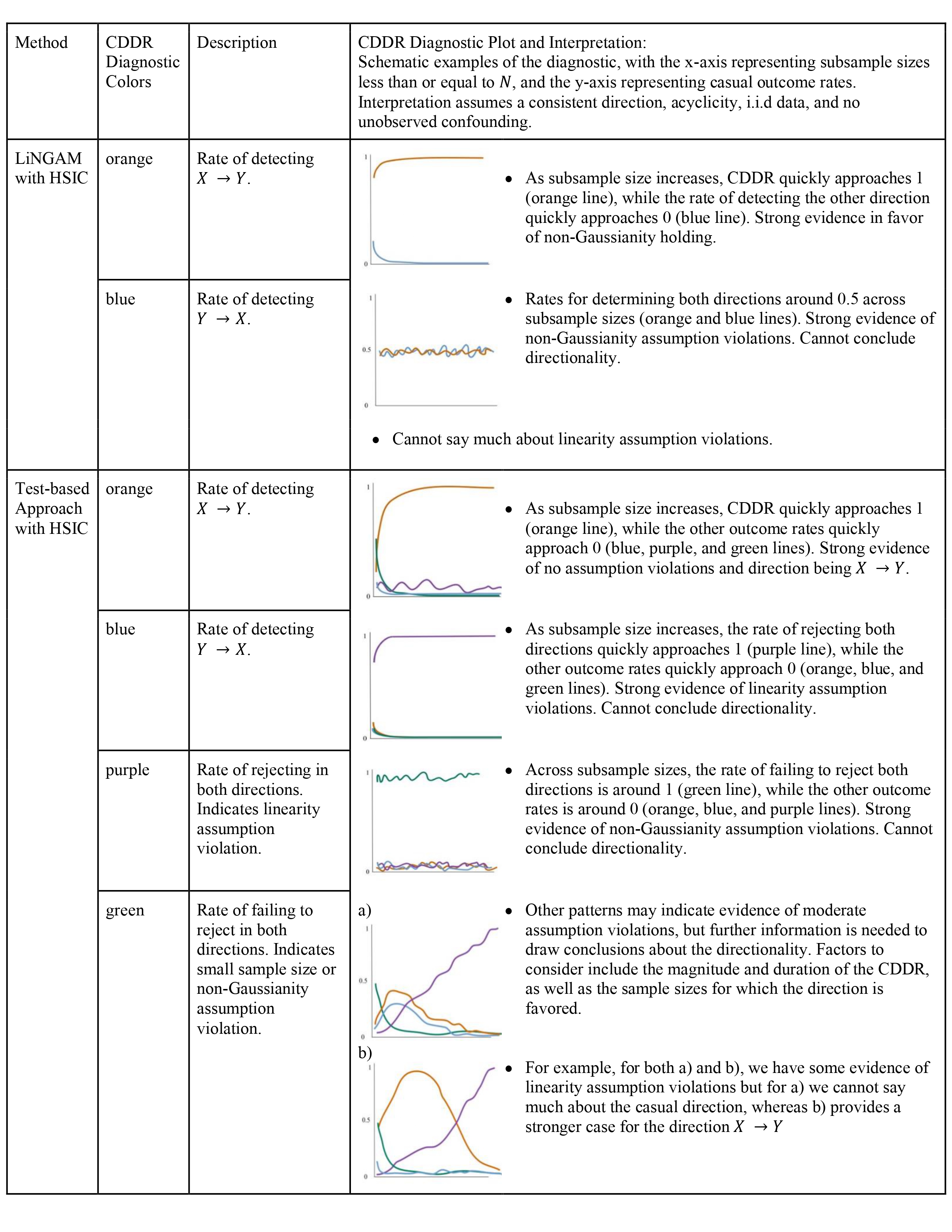}
    \label{Tab:sim_setup}
\end{table}

\begin{table}[]
    \footnotesize
    \centering
    \caption{Simulation Settings for Varying Linearity}
    \begin{tabular}{p{3cm}|p{3cm}|p{6cm}}
    \hline
        Linear & Polynomial = 1 & $Y = sign(X-a)|X-a|*\beta + \epsilon$\\ \hline
        Slightly Nonlinear & Polynomial = 1.25 & $Y = sign(X-a)|X-a|^{1.25}*\beta + \epsilon$\\ \hline
        Nonlinear & Polynomial = 3  & $Y = sign(X-a)|X-a|^3*\beta + \epsilon$\\
        \hline
    \end{tabular}
    \label{tab:lin_set_table}
\end{table}

\begin{table}[]
    \centering
    \caption{Simulation Settings for Varying Levels of Gaussianity, k is the number of mixtures}
    \begin{tabular}{p{4cm}|p{6cm}}
    \hline
        Gaussian & $X \sim N(0,1), \epsilon \sim N(\mu_1, \sigma_1)$\\ \hline
        Slightly Non-Gaussian & $X \sim N(0,1), \epsilon \sim GMM(k=2)$\\ \hline
        Non-Gaussian & $X \sim N(0,1), \epsilon \sim GMM(k=3)$ \\
        \hline
    \end{tabular}
    \label{tab:gauss_set_table}
\end{table}

We perform simulations with two dataset sizes: $N = 10,000$ and $N = 400$. In each simulation, we randomly sample data with replacement from the dataset of size $N$. For the larger dataset, the subsamples range from $20$ to $1699$ samples\footnote{We choose 20 as the smallest subsample size to represent a small, yet informative subsample size to determine the directionality. We choose 1699 as the largest subsample size to ensure a subsample size that is large but not too large to be computationally burdensome.}, and for the smaller dataset, the subsamples range from $20$ to $400$ samples. The results for $N = 400$ are included in the appendix, as they are similar to the results for $N = 10,000$.

\begin{figure}
     \centering
     \begin{subfigure}[b]{0.3\textwidth}
         \centering
         \includegraphics[width=\textwidth]{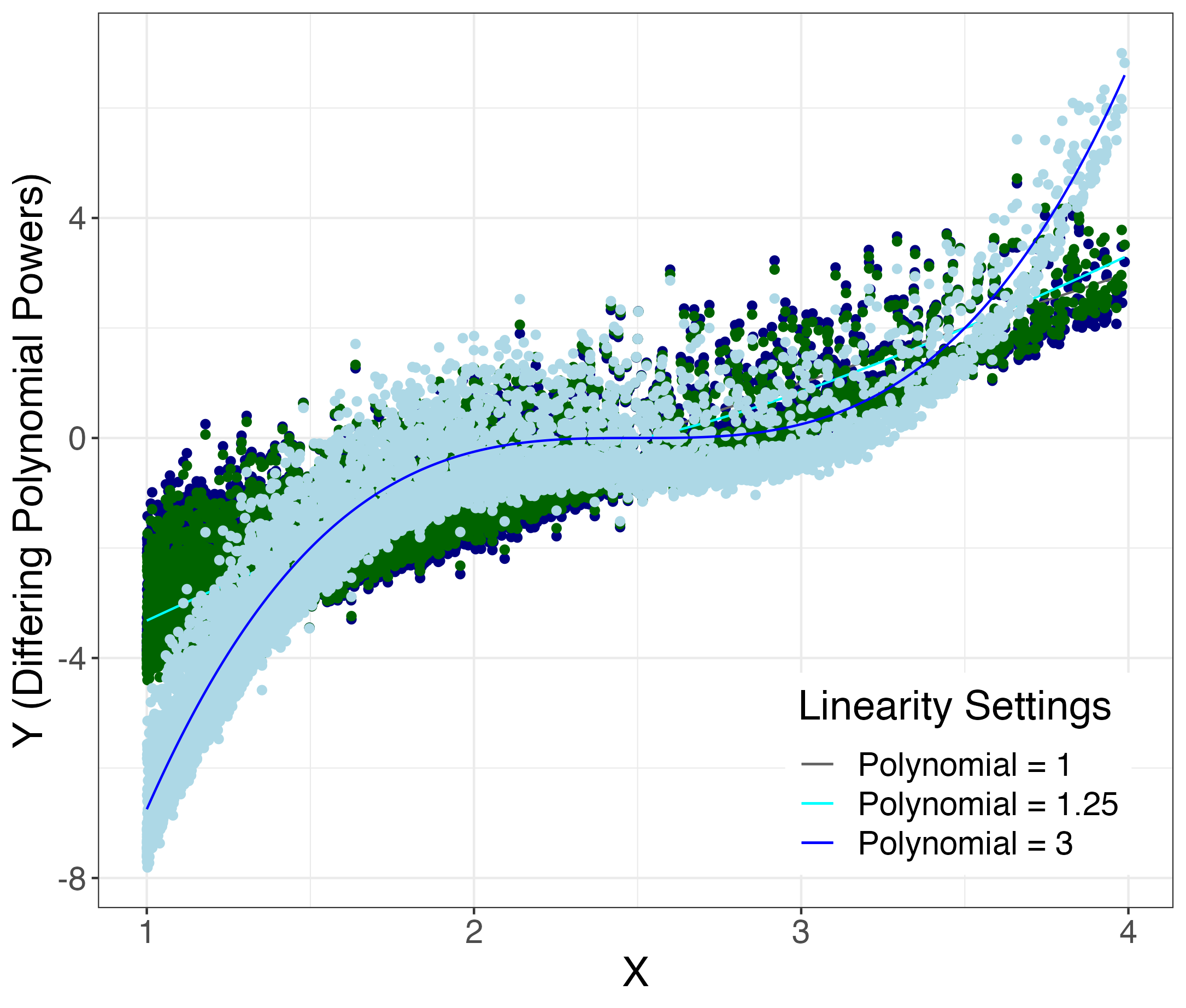}
        \caption{Settings of Linearity}
         \label{fig:lin}
     \end{subfigure}
     \hfill
     \begin{subfigure}[b]{0.3\textwidth}
         \centering
         \includegraphics[width=\textwidth]{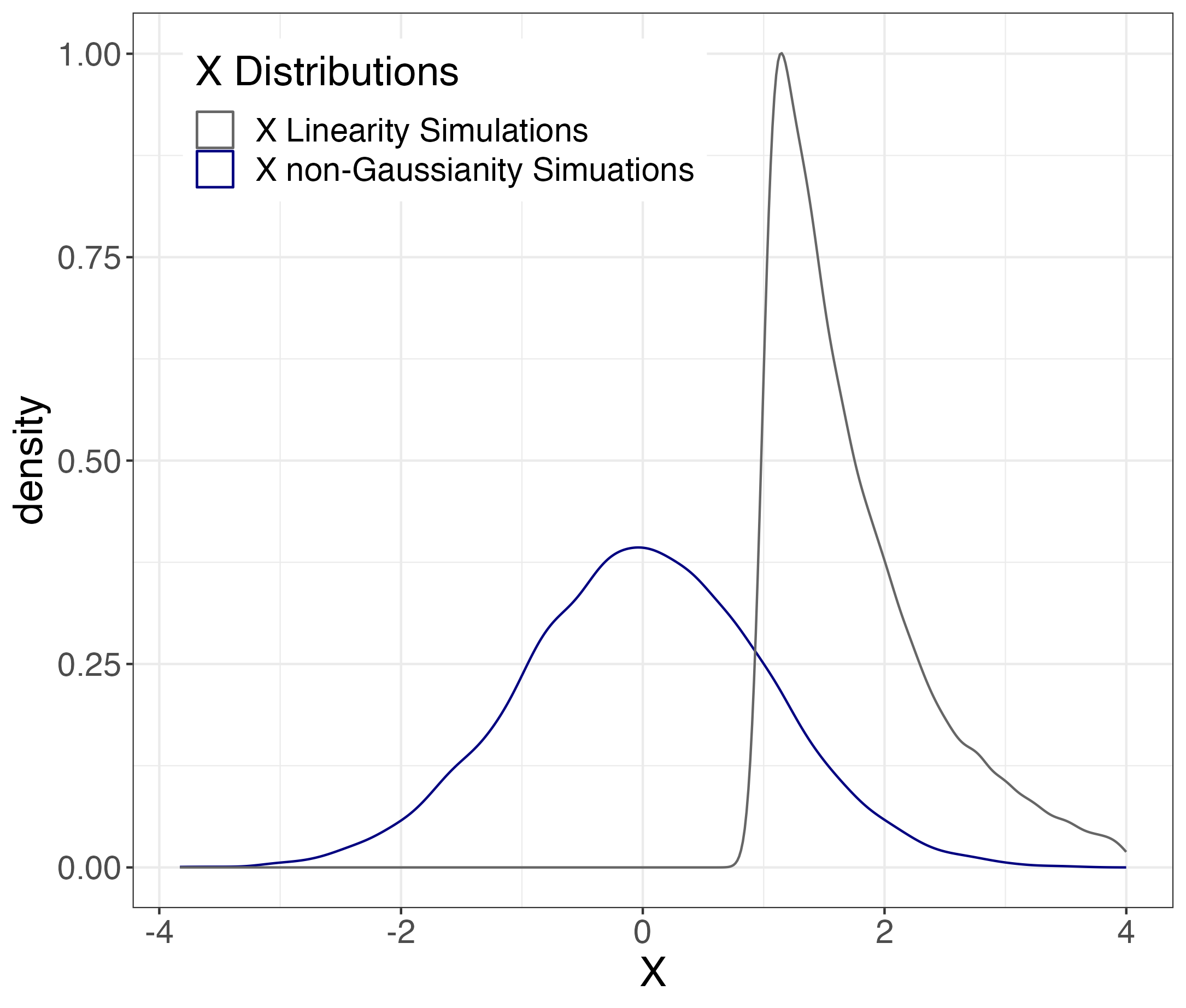}
         \caption{Simulation $X$ Distribution}
         \label{fig:gd}
     \end{subfigure}
     \hfill
     \begin{subfigure}[b]{0.3\textwidth}
         \centering
         \includegraphics[width=\textwidth]{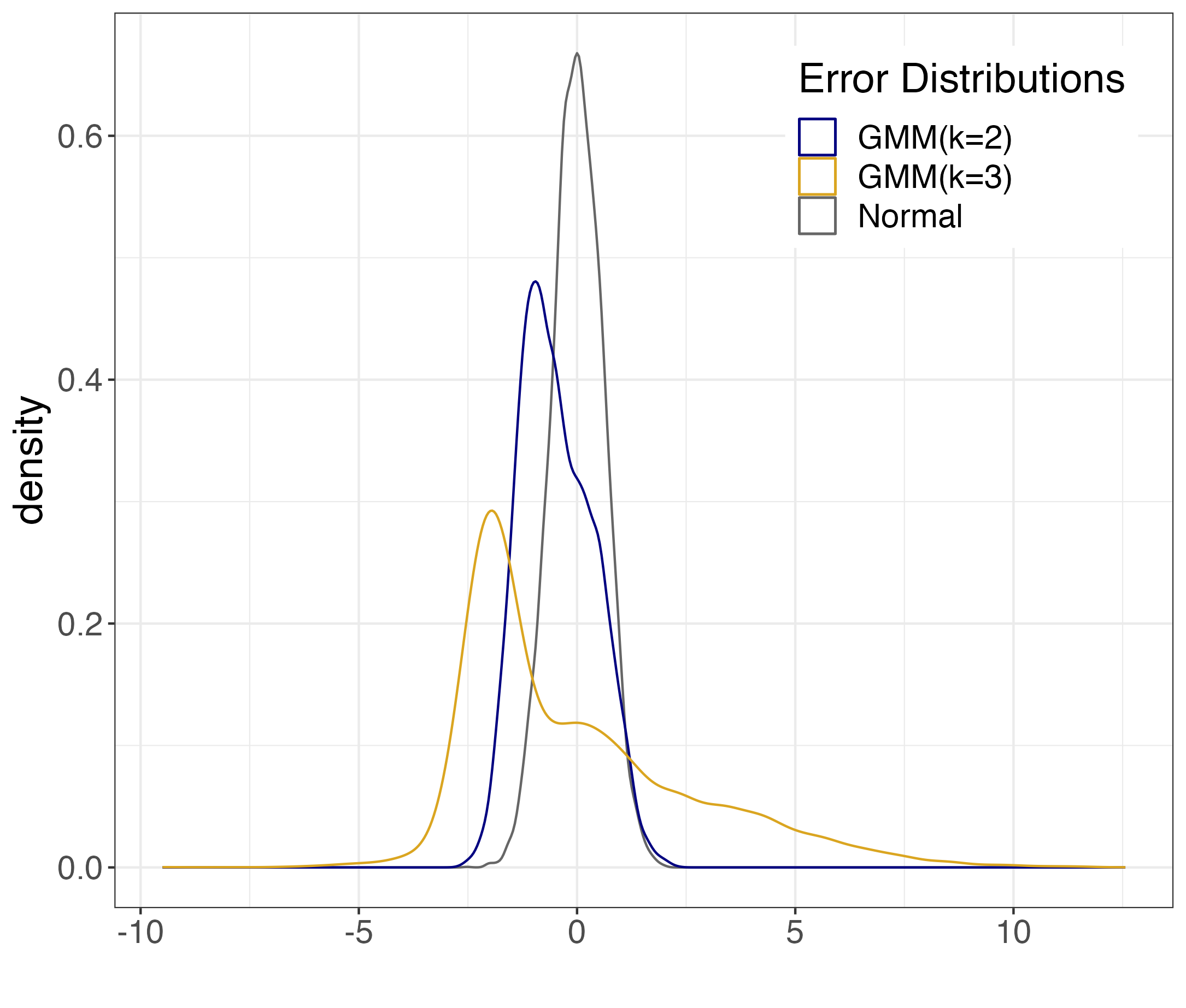}
         \caption{Settings of Gaussianity Errors}
         \label{fig:gerr}
     \end{subfigure}
        \caption{Settings of Simulations}
        \label{fig:three_graphs}
\end{figure}

In our simulations, the true direction is $X \rightarrow Y$. Table \ref{tab:lin_set_table} shows the three levels of linearity assumption violations considered: no violations (where $Y$ is a linear function of $X$), slight violations (where $Y$ is a polynomial function of $X$ with a polynomial power of 1.25), and strong violations (where $Y$ is a polynomial function of $X$ with a polynomial power of 3). We ensure all other assumptions, besides linearity, are met. We sample $X$ from a truncated exponential distribution (Figure \ref{fig:gd}), which ensures that $Y$ is not normally distributed. Figure \ref{fig:lin} illustrates the simulated data under each linearity setting.

Table \ref{tab:gauss_set_table} outlines the three levels of increasing Gaussianity, or degree of non-Gaussianity assumption violation, considered: non-Gaussianity holds (errors from a Gaussian Mixture Model (GMM) with 3 mixtures), slight violations (errors from a GMM with 2 mixtures), and severe violations (Gaussian data and errors). We ensure all other assumptions, besides non-Gaussianity, are met. We sample $X$ from a Gaussian distribution (Figure \ref{fig:gd}), such that if both $\epsilon$ and $X$ are Gaussian, $Y$ will also be Gaussian. Figure \ref{fig:gerr} illustrates the simulated errors under each non-Gaussianity setting. 

In Table \ref{Tab:sim_setup}, we explain how to read and interpret CDDR diagnostic for LiNGAM and the test-based causal discovery approaches under possible violations of linearity and non-Gaussianity. This interpretation assumes a consistent direction exists and that all other assumptions hold. In our simulations, we compare the DirectLiNGAM algorithm\footnote{We choose the DirectLiNGAM algorithm over the Independent Component Analysis (ICA)-based implementation of LiNGAM because DirectLiNGAM, when all necessary assumptions are met, is guaranteed to converge to the correct solution in a fixed number of iterations as the sample size approaches infinity \cite{shimizu2011directlingam}. This guarantee does not hold for the ICA-based version. In the bivariate context, DirectLiNGAM fits two linear models represented by Equations (\ref{eq:lingam_xy}) and (\ref{eq:lingam_yx}) and determines the causal direction by assessing the independence between the predictor and the residual. For both directions, the algorithm calculates a user-specified measure of independence and selects the direction that minimizes this measure.} using the Hilbert-Schmidt Independence Criterion (HSIC) with the test-based approach using the independence and goodness-of-fit test introduced by Sen and Sen \cite{sen_testing_2014}. Since the test-based approach with the Sen and Sen test uses the HSIC, we use the same independence measure for DirectLiNGAM to improve the comparability. For LiNGAM, CDDR diagnostic consists of two lines corresponding to the two possible causal discovery outcomes. The orange line is the rate at which we favor the correct direction $X \rightarrow Y$ for each subsample size, while the blue line is the rate at which we favor the incorrect direction $Y \rightarrow X$ for each subsample size. On the other hand, CDDR diagnostic for the test-based approach consists of four lines, one for each possible causal discovery outcome. Like CDDR diagnostic for LiNGAM, the orange line is the rate of favoring the correct direction $X \rightarrow Y$ for each subsample size, and the blue line is the rate of favoring the incorrect direction $Y \rightarrow X$ for each subsample size. The purple line is the rate of rejecting in both directions. When the linearity assumption is violated, we expect the rate of rejecting in both directions (purple line) to approach 1 as the subsample size increases. The green line is the rate of failing to reject in both directions. When the non-Gaussianity assumption is violated, we expect the rate of failing to reject in both directions (green line) to be around 1 across subsample sizes, as the causal direction is unidentifiable. Similarly, when the sample size is too small, there is insufficient signal to reject either hypothesis, so we would also expect the rate of failing to reject both directions (green line) to be high. 

In general, we expect the rate of detecting the correct direction ($X \rightarrow Y$) to decrease as the severity of assumption violations increases. It is unlikely for both linearity and non-Gaussianity assumptions to be violated simultaneously, as non-linear transformations of Gaussian distributed random variables are non-Gaussian. Overall, CDDR diagnostic helps visualize the interaction between assumption violations and sample size, which influences the ability of LiNGAM and the test-based approach to determine the correct causal direction. When applied to the test-based approach, CDDR diagnostic also provides insights into the extent of assumption violations.

\subsection{Linearity Simulation Results}
\label{subsec:lin_sims}

\begin{figure}
    \centering
    \includegraphics[width=190mm]{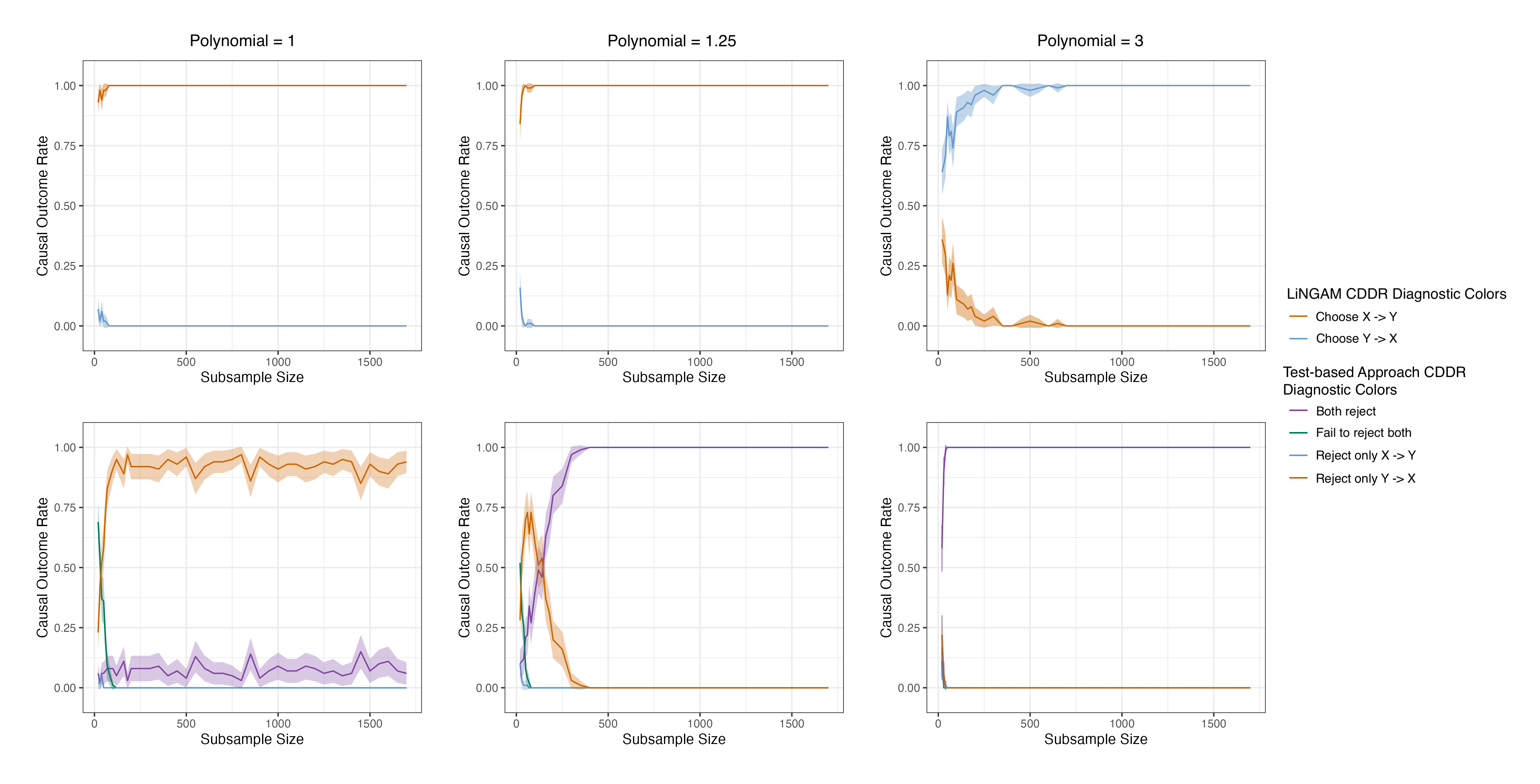}
    \caption{CDDR diagnostic for varying levels of linearity assumption violations. In each plot, the x-axis represents the subsample sizes, which are less than or equal to $N$, and the y-axis represents the causal outcome rates. The first row is CDDR diagnostic for LiNGAM. The second row is CDDR diagnostic for the test-based approach. The first column corresponds to no violations ($Y$ is a linear function of $X$), the second column corresponds to slight violations ($Y$ is a polynomial function of $X$ with a polynomial power of 1.25), and the third column corresponds to severe violations ($Y$ is a polynomial function of $X$ with a polynomial power of 3). The confidence regions correspond to pointwise confidence intervals.}
    \label{fig:sim_lins}
\end{figure}

Figure \ref{fig:sim_lins} shows CDDR diagnostic for varying levels of linearity assumption violations. For LiNGAM, CDDR diagnostic presents a consistent pattern in the absence of or with minor linearity violations: the rate of detecting the correct direction (orange line) quickly approaches 1 as subsample size increases, indicating that the correct direction ($X \rightarrow Y$) is favored. However, with severe violations, we observe a similar pattern but with the blue line rather than the orange line, signaling that in this case, LiNGAM would output the incorrect direction ($Y \rightarrow X$) due to assumption violations. Since the general pattern remains consistent regardless of the degree of violation, we find that CDDR diagnostic does not provide significant insights into the presence or severity of linearity assumption violations when applied to LiNGAM. 

For the test-based approach, when there are no violations, the rate of detecting the correct direction (orange line) quickly approaches 1 as the subsample size increases, while the other outcome rates quickly approach 0 (blue, purple, and green lines). Here, CDDR diagnostic provides strong evidence in favor of linearity holding and the directionality being $X$ causes $Y$ (correct direction). 

When there are slight assumption violations, the rate of detecting the correct direction (orange line) in our simulation study peaks at around 0.75 at approximately subsample size 100. As the subsample size increases beyond this peak, the test-based approach has enough samples to detect assumption violations with increasing strength, causing the rate of detecting the correct direction (orange line) to approach 0 while the rate of rejecting in both directions (purple line) approaches 1. CDDR diagnostic provides evidence of some linearity assumption violations. Additionally, the rate of detecting the correct direction is the dominant outcome rate for subsamples ranging from approximately 30 to 180. Given the magnitude, duration, and confidence intervals observed for this region, CDDR diagnostic supports $X \rightarrow Y$ outcome (correct direction), which may be useful for generating scientific hypotheses, although further investigation is needed to determine the directionality. Some suggestions for further exploration are given in some of our real data examples.

When there are severe assumption violations, we observe that the rate of rejecting in both directions (purple line) quickly approaches 1 as subsample size increases, while the other outcome rates (orange, blue, and green lines) quickly approach 0. Here, CDDR diagnostic provides strong evidence in favor of severe linearity assumption violations. Thus, we cannot conclude the directionality. 

In summary, for LiNGAM, CDDR diagnostic does not provide much information about linearity assumption violations. In contrast, for the test-based approach, CDDR diagnostic provides information about the existence and extent of linearity assumption violations as a function of sample size while still indicating the favored directionality, if one exists.

\subsection{Non-Gaussian Error Simulation Results}
\label{subsec:non_gauss_sims}

\begin{figure}
    \centering
    \includegraphics[width=190mm]{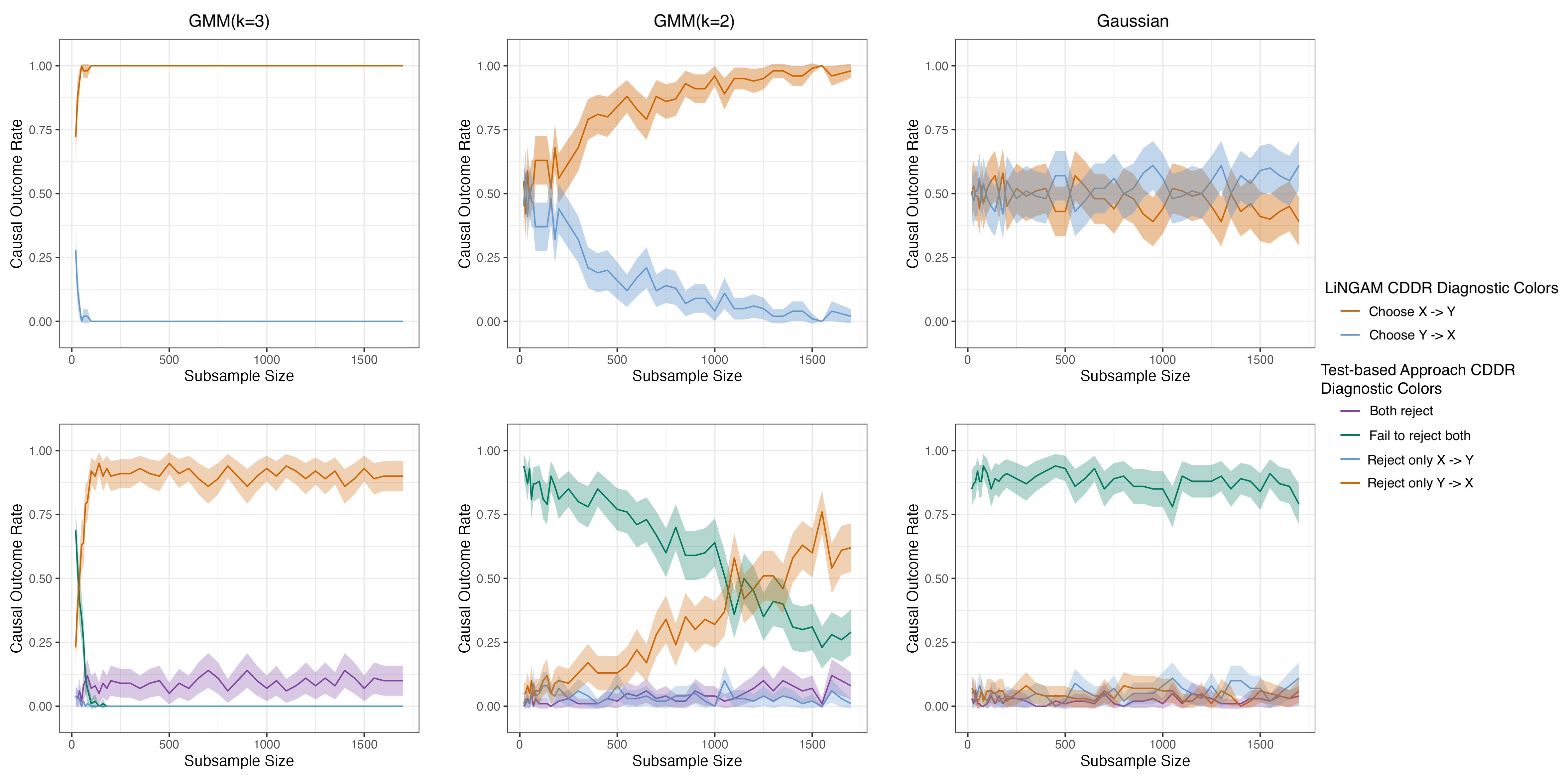}
    \caption{CDDR diagnostic for varying levels of non-Gaussianity assumption violations. In each plot, the x-axis represents the subsample sizes, which are less than or equal to $N$, and the y-axis represents the causal outcome rates. The first row is CDDR diagnostic for LiNGAM. The second row is CDDR diagnostic for the test-based approach. The first column corresponds to no violations (errors from a GMM(k=3), where k=3 signifies 3 mixtures), the second column corresponds to slight violations (errors from a GMM(k=2)), and the third column corresponds to severe violations (Gaussian data and errors). The confidence regions correspond to pointwise confidence intervals.}
    \label{fig:sim_gauss}
\end{figure}

Figure \ref{fig:sim_gauss} shows CDDR diagnostic for varying levels of non-Gaussianity assumption violations. In our simulations, for both LiNGAM and the test-based approach, when there are no assumption violations, the rate of detecting the correct direction (orange line) quickly approaches 1 as the subsample size increases, while the other outcome rate(s) quickly approaches 0. Here, CDDR diagnostic provides strong evidence for non-Gaussianity holding. Additionally, for the test-based approach, CDDR diagnostic provides strong evidence supporting the direction $X$ causes $Y$ (correct direction). For LiNGAM, CDDR diagnostic favors this direction, but it is crucial to verify that all other assumptions are met before drawing any conclusions.

For LiNGAM, when there are slight assumption violations, the rate of detecting the correct direction (orange line) slowly increases from around $0.5$ to  $1$ as the subsample size grows. In contrast, the rate of determining the incorrect direction (blue line) slowly decreases from around $0.5$ to $0$ as the subsample size increases. For the test-based approach, the rate of failing to reject both directions (green line) gradually approaches 0 as the subsample size increases. The rate of detecting the correct direction (orange line) slowly grows to a rate higher than $0.5$ as the subsample size increases. All other outcome rates remain around $0$ across subsample sizes. For both LiNGAM and the test-based approach, CDDR diagnostic provides evidence of slight non-Gaussianity assumption violations. CDDR diagnostic also indicates that both methods favor the correct direction ($X \rightarrow Y$). For LiNGAM, assuming all other assumptions are met, this provides substantial evidence supporting the direction $X \rightarrow Y$ (correct direction). For the test-based approach, the rate of detecting the correct direction is the dominant outcome rate for subsamples ranging from approximately $1200$ to $1700$. Given the magnitude, duration, and confidence intervals observed for this region, CDDR diagnostic supports $X \rightarrow Y$ (correct direction), but further investigation is needed to determine the directionality. 

For LiNGAM, when there are severe assumption violations, the rates of determining both directions are around $0.5$ across subsample sizes (orange and blue lines). For the test-based approach, the rate of failing to reject in both directions (green line) quickly approaches 1 as the subsample size increases, while the other outcome rates (orange, blue, and purple lines) quickly approach 0. For both methods, CDDR diagnostic provides strong evidence in favor of non-Gaussianity assumption violations. Thus, we cannot conclude the directionality. 

In summary, our simulation studies demonstrate that for both LiNGAM and the test-based approach, CDDR diagnostic provides information about the existence and extent of non-Gaussianity assumption violations as a function of subsample size while still indicating the favored directionality, if one exists.

\section{Real Data}
\label{sec:real_data}

We demonstrate the use of CDDR diagnostic with LiNGAM and the test-based causal discovery approach on real-world datasets including two cause-effect benchmark  T\"{u}bengen pairs, the Ozone and Temperature and the Population and Food Consumption Data~\cite{mooij_distinguishing_2016}. In Section \ref{subsec:ozone}, we discuss results for the Ozone and Temperature dataset. In Section \ref{subsec:food_data}, we present results for the Population and Food Consumption Dataset. In Section \ref{subsec:trout}, we analyze the Rainbow Trout dataset, obtained from the Organization for Economic Co-operation and Development (OECD)~\cite{oecd2006current} and previously analyzed using the Dose-Response Curve R package by Ritz et al~\cite{ritz2015dose}. These three datasets have a known causal direction. Finally, in Section \ref{subsec:sleep_depr}, we analyze the Sleep and Depression data from a causal discovery study by Rosenstr\"{o}m et al.~\cite{rosenstrom2012pairwise}, where the causal direction is unknown.

\subsection{Ozone and Temperature Data}
\label{subsec:ozone}

The Ozone and Temperature dataset contains 365 daily mean temperatures and lower atmosphere ozone concentrations from Davos-See, Switzerland in 2009. The data is collected by the Swiss Federal Office for the Environment (FOEN) \cite{adminAirData}. Following Mooji et al.~\cite{mooij_distinguishing_2016}, we assume the causal direction is known with changes in temperature causing changes in lower atmosphere ozone concentrations. When plotting ozone concentrations against temperature in Figure \ref{fig:real_res}, it appears that the relationship between the two variables is approximately linear. When interpreting CDDR diagnostic (following the guidelines in Table \ref{Tab:sim_setup}), we assume that all other assumptions hold and that there exists a causal direction. 
 
The top middle and right plots in Figure \ref{fig:real_res} present CDDR diagnostics for LiNGAM and the test-based approach for this dataset. We observe that LiNGAM is confident in selecting the incorrect direction for all subsamples. On the other hand, CDDR diagnostic paired with the test-based approach indicates moderate to severe linearity assumption violations and supports the inconclusive causal outcome. While this is the overall conclusion, when we focus on the two deterministic directions, we see that CDDR diagnostic paired with the test-based approach favors the hypothesized direction. This is because the test-based approach calculates p-values, which compare the test statistics to the corresponding null distributions, instead of solely comparing test statistics as LiNGAM does.

Here, we see an example where simply running LiNGAM on the entire dataset would result in picking the incorrect direction as the causal discovery outcome. In contrast, using CDDR diagnostic alongside the test-based approach detects subtle non-linearity in the real data. These subtle non-linearities are nonetheless severe enough to affect causal discovery, supporting the inconclusive outcome.

\subsection{Population and Food Consumption Data}
\label{subsec:food_data}
\begin{figure}
    \centering
    \includegraphics[width=192mm]{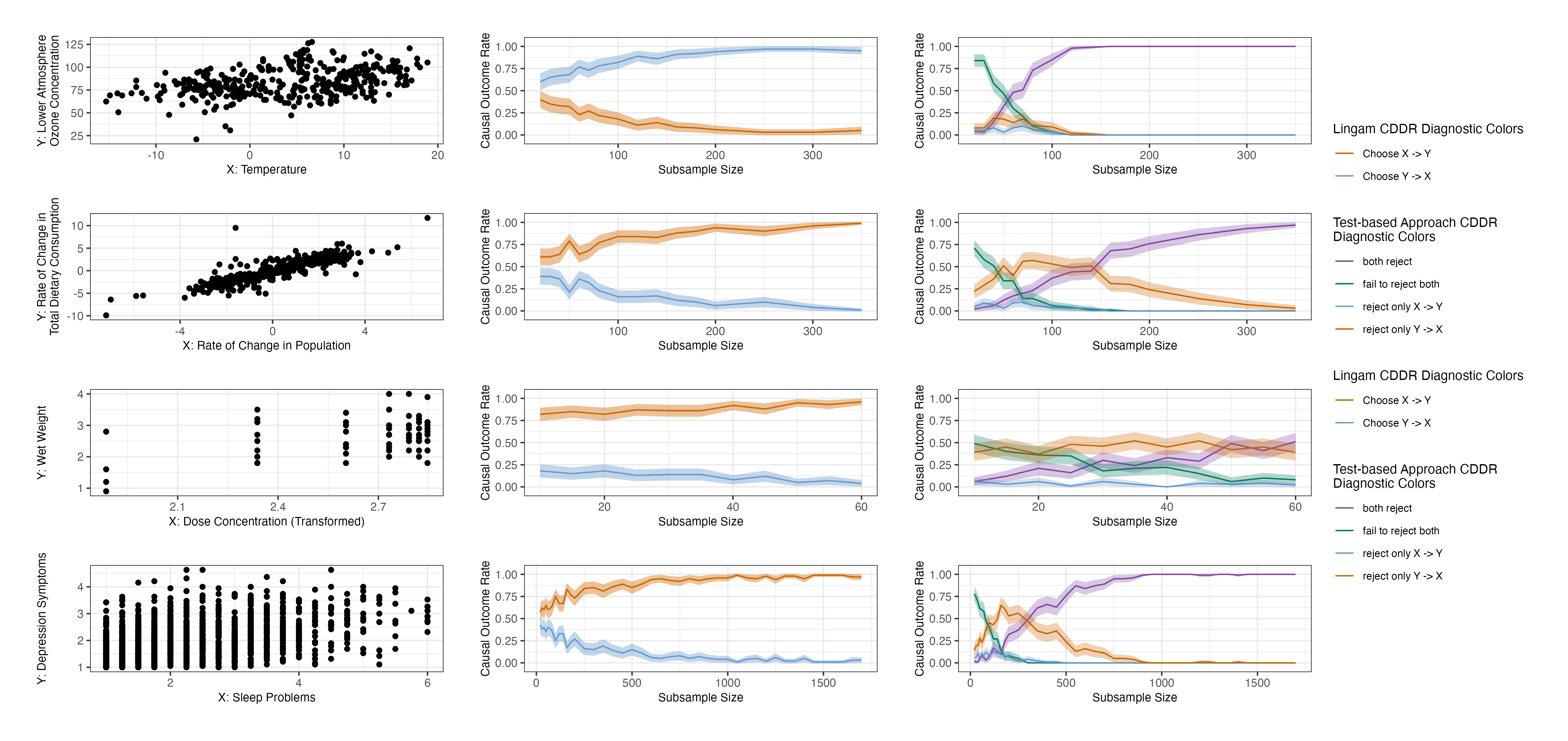}
    \caption{Real Data Results. The first row represents the Ozone and Temperature dataset, the second row represents the Population and Dietary Consumption Dataset, the third row represents the Rainbow Trout dataset, and the fourth row represents the Sleep and Depression dataset. The first column is the scatter plot of the data. The second column is the LiNGAM results, and the third column represents the test-based approach results.}
    \label{fig:real_res}
\end{figure}  

The Population and Food Consumption dataset contains 347 data points from 174 countries or areas. It covers the years from 1990-92 to 1995-97 (former 174 data points) and from 1995-97 to 2000-02 (latter 174 points), with one missing entry. The data is collected by the Food and Agriculture Organization of the United Nations. This dataset records the average annual rate of change in population alongside the average annual rate of change in total dietary consumption. Following Mooji et al.~\cite{mooij_distinguishing_2016}, we assume that the hypothesized (known) causal direction is that population change causes a change in total dietary consumption. From the scatterplot in Figure \ref{fig:real_res}, it appears that the relationship between the two variables is approximately linear. In our interpretation of CDDR diagnostic (following the guidelines in Table \ref{Tab:sim_setup}), we assume that all other assumptions approximately hold and that there is a consistent direction. 

The second row in Figure \ref{fig:real_res} presents CDDR diagnostic applied to both LiNGAM and the test-based approach for this dataset. CDDR diagnostic indicates that both methods favor the hypothesized direction, with CDDR diagnostic for the test-based approach providing some evidence of linearity assumption violations. The rate of detecting the hypothesized direction is the dominant outcome rate for subsamples ranging from approximately 50 to 150. Given the magnitude, duration, and confidence intervals observed for this region, CDDR diagnostic provides some support for the hypothesized causal direction, which may be useful for generating scientific hypotheses. However, further investigation is needed to determine the directionality. This may include choosing a more appropriate method with fewer assumption violations to determine directionality, if it exists. Alternatively, external information, such as prior beliefs, can be utilized along with the information given by CDDR diagnostic to estimate the directionality. Additionally, collecting more informative data, such as sequential or experimental data, can also help in learning the directionality.

Here, we see an example where there are slight to moderate linearity assumption violations, but CDDR diagnostic applied to both LiNGAM and the test-based approach provides some support for the known direction for moderate sample sizes, however at larger sample sizes, the test-based approach picks up on the slight non-linearities in the data.

\subsection{Rainbow Trout Data}
\label{subsec:trout}

The Rainbow Trout dataset describes data from a 21-day fish test \cite{oecd2006current}. It uses the test organism Rainbow trout or Oncorhynchus mykiss. The trouts were held in 14-15 \degree C water and exposed to one of 7 concentrations (6 non-zero concentrations and one control) of an unknown agent. After 28 days, the wet weight was recorded \cite{jensen2020bmd}. The known causal direction is that changes in dose concentration cause changes in wet weight; this is our hypothesized direction based on the experimental design. The weights were recorded for each concentration 10 times, however, there were some missing values at higher doses, resulting in the total of 61 observations. Dose-response data are well studied and known to follow a particular structure \cite{murado2002dose}. It is established in OECD~\cite{oecd2006current} to fit a two-parameter exponential decay to the Rainbow trout data. We use this established dose-response model from pharmacology to transform the data into an approximately linear form. Then, we apply LiNGAM and the test-based approach to the transformed data. For details on applying LiNGAM and the test-based approach to scientifically derived data transformations, please see the appendix. The transformed data in Figure \ref{fig:real_res} shows that the relationship between the two variables is approximately linear. Our interpretation of CDDR diagnostic (following the guidelines in Table \ref{Tab:sim_setup}) assumes a consistent direction and that all other assumptions approximately hold. As this is a randomized experiment this is likely true by design.

The third row right and middle plots in Figure \ref{fig:real_res} present CDDR diagnostic applied to LiNGAM and the test-based approach for this dataset. CDDR diagnostic indicates that both methods favor the hypothesized direction. CDDR diagnostic for the test-based approach detects the non-linearities of the data for larger subsample sizes (around 50 to 60). Thus, larger sample sizes allow the test-based approach to pick up on the slight non-linearities. The rate of detecting the hypothesized direction is the dominant outcome rate for subsamples ranging from approximately 15 to 45. Given the magnitude, duration, and confidence intervals observed for this region, CDDR diagnostic supports the direction change in dose causes a change in wet weight. Here, we see an example where one can use CDDR diagnostic to provide support for their directionality estimate, as it detects minor linearity assumption violations. 

\subsection{Sleep and Depression Data}
\label{subsec:sleep_depr}

Rosenstr\"{o}m et al.~\cite{rosenstrom2012pairwise} used LiNGAM on this dataset to study directionality of the causal relationship between sleep problems and depression. 
While sleep problems often precede depression symptoms, they can also be a prodromal sign of depression. Therefore, the temporal order would not be enough to determine the causal relationship. This dataset has 1699 observations on Finnish residents' (aged 30 to 45) sleep and depression scores. The survey was given out to randomly sampled residents based on their social security numbers. Only fully completed responses are included in the dataset. Rosenstr\"{o}m et al.~\cite{rosenstrom2012pairwise} ran LiNGAM on the dataset and found evidence that sleep problems caused depressive symptoms; this will be our hypothesized direction. The sleep problems and depression symptoms scatterplot in Figure \ref{fig:real_res} indicated an approximately linear relationship between the two variables. Our interpretation of CDDR diagnostic (following the guidelines in Table \ref{Tab:sim_setup}) assumes that other assumptions approximately hold and that there is a consistent direction. Rosenstr\"{o}m et al.~\cite{rosenstrom2012pairwise} gave reasons in their paper\footnote{Rosenstr\"{o}m et al.~\cite{rosenstrom2012pairwise} state that ``recent studies suggest that depressive symptoms form a causal network of symptoms that directly influence each other, instead of reflecting a single latent causal antecedent. This suggests that the association between sleep problem symptoms and other depressive symptoms is not fully confounded by a latent third factor, but a detectable dominant causal direction may exist''. Additionally, the data is approximately i.i.d as the participants are randomly contacted.} in support of our assumptions.

The middle and right plots in the bottom row in Figure \ref{fig:real_res} presents CDDR diagnostic applied to both LiNGAM and the test-based approach for this dataset which indicates that both methods favor the correct direction. However, CDDR diagnostic for the test-based approach provides evidence of linearity assumption violations. The rate of detecting the hypothesized direction is the dominant outcome rate for subsamples ranging from approximately 80 to 300. Given the magnitude, duration, and confidence intervals observed for this region, CDDR diagnostic provides some support for the direction sleep problems cause depressive symptoms, which is the outcome reported by Rosenstr\"{o}m et al.~\cite{rosenstrom2012pairwise}. However, further investigation is needed to determine the directionality. Again this may include using other methods, prior beliefs, or more informative data sources to learn the directionality if one exists.

Here, we apply CDDR diagnostic to an example where previously only LiNGAM, without any causal discovery diagnostic, was used. With CDDR diagnostic we see that there are likely assumption violations but it also provides some support in favor of the directionality previously reported by Rosenstr\"{o}m et al.~\cite{rosenstrom2012pairwise}.

\section{Discussion}

We have proposed CDDR diagnostic: a diagnostic tool for bivariate causal discovery to evaluate how the interplay between assumption violations and sample size affects the estimated causal directionality. We demonstrated its use in detecting linearity and non-Gaussianity assumption violations for two methods: LiNGAM and our test-based approach. In our simulations and real data analyses, we found CDDR diagnostic to be an effective tool, especially when used alongside a causal discovery method that offers more information than just a deterministic direction. For instance, when applied to LiNGAM, CDDR diagnostic could reveal the effect of non-Gaussianity assumption violations on causal discovery. In addition, when applied to the more informative test-based approach, it reveals how the presence and extent of linearity and non-Gaussianity assumption violations affect causal discovery.

We demonstrated CDDR diagnostic for two methods in the linear setting. However, this approach can be applied to any bivariate method that uses the independence between the noise and predictor to identify the directionality, such as bivariate functional causal discovery methods \cite{pmlr-v206-keropyan23a, shimizu2006linear, shimizu2011directlingam, hoyer2009nonlinear, zhang2012identifiability}. In this generalized setting, CDDR diagnostic evaluates the goodness-of-fit of the functional causal model rather than linearity, and checks for identifiability issues rather than non-Gaussianity assumption violations. To apply CDDR diagnostic to, for example, Additive Noise Models (ANMs), one can simply follow the same procedure outlined in Algorithm \ref{alg:dir_rate} and plug in an ANM method as the causal discovery method of choice. However, to fully benefit from CDDR diagnostic, one would need to pair the diagnostic with a test-based version of ANMs, which currently does not exist. One can easily use our paper and \R{} package, \href{https://github.com/shreyap18/causalDiagnose}{CasualDiagnose}, to extend our test-based approach to ANMs. To do this, one can input the nonlinear functions used in both directions for the ANM model into our package. For example, if the ANM corresponding to each direction is expressed as $Y = f_{AN}(X) + \epsilon$ for $X \rightarrow Y$ and $X = g_{AN}(Y) + \eta$ for $Y \rightarrow X$, one would need to provide the estimated generating functions $\hat{f}_{AN}$ and $\hat{g}_{AN}$. Our package will then estimate p-values corresponding to the pair of hypothesis tests in Equation (\ref{eq:h_indep}). It currently estimates the p-values for the Sen and Sen goodness-of-fit and independence test~\cite{sen_testing_2014} using bootstrap, but we plan to add support for other selected independence tests in the future. Our package then compares these p-value estimates to a significance level to determine the outcome of the test-based approach (e.g., reject $Y\rightarrow X$ and fail to reject $X\rightarrow Y$). One could then use our package to plot CDDR diagnostic applied to such a test-based approach for ANMs.

There is a paucity of recent work aimed at quantifying uncertainty in causal discovery \cite{chang2024post, wang2023confidence,gradu2022valid} by using ``confidence sets'' of the causal orderings to measure uncertainty. Out of these methods, Wang et al.~\cite{wang2023confidence} is the only method applicable to functional causal discovery. Wang et al.~\cite{wang2023confidence} construct a confidence set of causal orderings by inverting a goodness-of-fit test. In the bivariate setting, these confidence sets are unlikely to be informative as there are only two causal orderings: $X \rightarrow Y$ and $Y \rightarrow X$. Additionally, their method provides limited information in the presence of assumption violations: if any assumptions are violated, it produces an empty confidence set. This overlooks the fact that inference on causal directionality might still be accurate, even in the presence of assumption violations. In contrast, our diagnostic tool is the first to study the effect of assumption violations on the causal directionality.

We demonstrated CDDR diagnostic for linearity and non-Gaussianity assumption violations as these assumptions are crucial for identifying the causal direction in the linear setting. However, other assumption violations, such as unobserved confounding, cyclicity, and non-i.i.d data, could affect the causal discovery outcomes as well. While CDDR diagnostic applied to LiNGAM can only detect violations that result in identifiability issues, CDDR diagnostic applied to the test-based approach can detect general assumption violations. Thus, when assumption violations result in identifiability issues, we expect the rate of failing to reject in both directions (green line) for the test-based causal discovery to approach 1. For other assumption violations (e.g., unobserved confounding, non-i.i.d data, etc.), we expect the rate of rejecting in both directions (purple line) to approach 1.

In this paper, we study CDDR diagnostic for the bivariate setting but it can be directly applied to multivariate cases. The simplest and most natural example would be if we knew the causal structure for all variables except for two variables of interest, $X$ and $Y$, and we understood how the other variables affect $X$ and $Y$. We could then use our methods to study the directionality between $X$ and $Y$ in this context with minor modifications. First, we would need to remove the effect of the known confounders on $X$ and $Y$ by modeling $X$ and $Y$, respectively, while controlling for these confounders (e.g., using linear regression). Then we could use the resulting residuals $r_X$ and $r_Y$ as a new pair of bivariate variables and apply our methods directly to them (i.e. estimating the possible causal discovery outcome rates as a function of sample size with $r_X$ and $r_Y$ as inputs to the causal discovery method).

One could extend CDDR diagnostic to a general multivariate setting where the causal structure among variables is not known. In that case, the number of causal discovery outcomes is likely to be large, thus, plotting all of these outcomes simultaneously as a function of subsample size may be confusing and computationally burdensome. Future work is required to determine how to summarize these outcomes in an informative and computationally efficient manner that still demonstrates the interplay between assumption violations and sample size. 

This paper takes the first step towards providing a general diagnostic tool for functional causal discovery. We encourage researchers to use CDDR diagnostic for careful evaluation of assumption violations and the potential impact of their sample size on causal discovery outcomes before drawing conclusions. 

\section*{Acknowledgments}
Shreya Prakash's work was partly supported by funding from the Center of Statistics and the Social Sciences (CSSS). The authors are grateful for feedback and support from the Statistical and Machine Learning Approaches for the Social Sciences working group as well as other faculty and Ph.D. students at the University of Washington.

\section*{Financial disclosure}

None reported.

\section*{Conflict of interest}

The authors declare no potential conflict of interests.

\bibliographystyle{plain}
\bibliography{causal-discovery}

\begin{thebibliography}{10}

\bibitem{chang2024post}
Ting-Hsuan Chang, Zijian Guo, and Daniel Malinsky.
\newblock Post-selection inference for causal effects after causal discovery.
\newblock {\em arXiv preprint arXiv:2405.06763}, 2024.

\bibitem{darmois_1953}
G~DARMOIS, F~DIVISIA, E~MORICE, J~RUEFF, G~MALECOT, EN~G{\'E}N{\'E}TIQUE LES MOD{\`E}LES~STOCHASTIQUES, CONCEPTION STOCHASTIQUE DE~COEFFICIENTS MULTIPLICATEURS, M~SKLAR, ET~LEURS MARGES, P~THIONET, et~al.
\newblock De l'universit{\'e} de paris.
\newblock {\em Publications de l'Institut de statistique de l'Universit{\'e} de Paris}, page 231, 1953.

\bibitem{EfroTibs93}
Bradley Efron and Robert~J. Tibshirani.
\newblock {\em An Introduction to the Bootstrap}.
\newblock Number~57 in Monographs on Statistics and Applied Probability. Chapman \& Hall/CRC, Boca Raton, Florida, USA, 1993.

\bibitem{adminAirData}
Federal~Office for~the Environment~FOEN.
\newblock {A}ir: {D}ata, indicators and maps --- bafu.admin.ch.
\newblock \url{www.bafu.admin.ch/luft/luftbelastung/blick_zurueck/datenabfrage/index.html?lang=de}, 2009.

\bibitem{glymour_review_2019}
Clark Glymour, Kun Zhang, and Peter Spirtes.
\newblock Review of {Causal} {Discovery} {Methods} {Based} on {Graphical} {Models}.
\newblock {\em Frontiers in Genetics}, 10, 2019.

\bibitem{gradu2022valid}
Paula Gradu, Tijana Zrnic, Yixin Wang, and Michael~I Jordan.
\newblock Valid inference after causal discovery.
\newblock {\em arXiv preprint arXiv:2208.05949}, 2022.

\bibitem{hoyer2009nonlinear}
Patrik~O. Hoyer, Dominik Janzing, Joris~M Mooij, Jonas Peters, and Bernhard Schölkopf.
\newblock Nonlinear causal discovery with additive noise models.
\newblock In D.~Koller, D.~Schuurmans, Y.~Bengio, and L.~Bottou, editors, {\em Advances in {Neural} {Information} {Processing} {Systems} 21}, pages 689--696. Curran Associates, Inc., 2009.

\bibitem{hu2018application}
Pengfei Hu, Rong Jiao, Li~Jin, and Momiao Xiong.
\newblock Application of causal inference to genomic analysis: advances in methodology.
\newblock {\em Frontiers in Genetics}, 9:238, 2018.

\bibitem{jensen2020bmd}
Signe~M Jensen, Felix~M Kluxen, Jens~C Streibig, Nina Cedergreen, and Christian Ritz.
\newblock bmd: an r package for benchmark dose estimation.
\newblock {\em PeerJ}, 8:e10557, 2020.

\bibitem{jiang2023signet}
Zhongli Jiang, Chen Chen, Zhenyu Xu, Xiaojian Wang, Min Zhang, and Dabao Zhang.
\newblock Signet: transcriptome-wide causal inference for gene regulatory networks.
\newblock {\em Scientific Reports}, 13(1):19371, 2023.

\bibitem{pmlr-v206-keropyan23a}
Grigor Keropyan, David Strieder, and Mathias Drton.
\newblock Rank-based causal discovery for post-nonlinear models.
\newblock In Francisco Ruiz, Jennifer Dy, and Jan-Willem van~de Meent, editors, {\em Proceedings of The 26th International Conference on Artificial Intelligence and Statistics}, volume 206 of {\em Proceedings of Machine Learning Research}, pages 7849--7870. PMLR, 25--27 Apr 2023.

\bibitem{kotoku2020causal}
Jun’ichi Kotoku, Asuka Oyama, Kanako Kitazumi, Hiroshi Toki, Akihiro Haga, Ryohei Yamamoto, Maki Shinzawa, Miyae Yamakawa, Sakiko Fukui, Keiichi Yamamoto, et~al.
\newblock Causal relations of health indices inferred statistically using the directlingam algorithm from big data of osaka prefecture health checkups.
\newblock {\em Plos one}, 15(12):e0243229, 2020.

\bibitem{mooij_distinguishing_2016}
Joris~M. Mooij, Jonas Peters, Dominik Janzing, Jakob Zscheischler, and Bernhard Schölkopf.
\newblock Distinguishing {Cause} from {Effect} {Using} {Observational} {Data}: {Methods} and {Benchmarks}.
\newblock {\em Journal of Machine Learning Research}, 17(32):1--102, 2016.

\bibitem{murado2002dose}
MA~Murado, Ma~P Gonz{\'a}lez, and JA~V{\'a}zquez.
\newblock Dose--response relationships: an overview, a generative model and its application to the verification of descriptive models.
\newblock {\em Enzyme and Microbial Technology}, 31(4):439--455, 2002.

\bibitem{oecd2006current}
OECD.
\newblock Current approaches in the statistical analysis of ecotoxicity data: A guidance to application, 2006.

\bibitem{peters_2008}
Jonas Peters.
\newblock {\em Asymmetries of time series under inverting their direction}.
\newblock PhD thesis, Universit{\"a}t Heidelberg Heidelberg, Germany, 2008.

\bibitem{ritz2015dose}
Christian Ritz, Florent Baty, Jens~C Streibig, and Daniel Gerhard.
\newblock Dose-response analysis using r.
\newblock {\em PloS one}, 10(12):e0146021, 2015.

\bibitem{rosenstrom2020distribution}
Tom Rosenstr{\"o}m and Regina Garc{\'\i}a-Vel{\'a}zquez.
\newblock Distribution-based causal inference.
\newblock {\em Direction dependence in statistical modeling: Methods of analysis}, page 267, 2020.

\bibitem{rosenstrom2012pairwise}
Tom Rosenstr{\"o}m, Markus Jokela, Sampsa Puttonen, Mirka Hintsanen, Laura Pulkki-R{\aa}back, Jorma~S Viikari, Olli~T Raitakari, and Liisa Keltikangas-J{\"a}rvinen.
\newblock Pairwise measures of causal direction in the epidemiology of sleep problems and depression.
\newblock {\em PloS one}, 7(11):e50841, 2012.

\bibitem{saito2023causal}
Tatsuki Saito and Koichi Fujiwara.
\newblock Causal analysis of nitrogen oxides emissions process in coal-fired power plant with lingam.
\newblock {\em Frontiers in Analytical Science}, 3:1045324, 2023.

\bibitem{sen_testing_2014}
A.~Sen and B.~Sen.
\newblock Testing independence and goodness-of-fit in linear models.
\newblock {\em Biometrika}, 101(4):927--942, December 2014.
\newblock Publisher: Oxford Academic.

\bibitem{shimizu2006linear}
Shohei Shimizu, Patrik~O Hoyer, Aapo Hyv{\"a}rinen, and Antti Kerminen.
\newblock A linear non-gaussian acyclic model for causal discovery.
\newblock {\em Journal of Machine Learning Research}, 7(Oct):2003--2030, 2006.

\bibitem{shimizu2011directlingam}
Shohei Shimizu, Takanori Inazumi, Yasuhiro Sogawa, Aapo Hyvarinen, Yoshinobu Kawahara, Takashi Washio, Patrik~O Hoyer, Kenneth Bollen, and Patrik Hoyer.
\newblock Directlingam: A direct method for learning a linear non-gaussian structural equation model.
\newblock {\em Journal of Machine Learning Research-JMLR}, 12(Apr):1225--1248, 2011.

\bibitem{shimizu_use_2008}
Shohei Shimizu and Yutaka Kano.
\newblock Use of non-normality in structural equation modeling: {Application} to direction of causation.
\newblock {\em Journal of Statistical Planning and Inference}, 138(11):3483--3491, 2008.

\bibitem{skitovic_1962}
VP~Skitovi{\v{c}} et~al.
\newblock Linear combinations of independent random variables and the normal distribution law.
\newblock 1962.

\bibitem{skitovic_1954}
Viktor~Pavlovich Skitovich.
\newblock Linear forms of independent random variables and the normal distribution law.
\newblock {\em Izvestiya Rossiiskoi Akademii Nauk. Seriya Matematicheskaya}, 18(2):185--200, 1954.

\bibitem{soloveychik2020central}
Ilya Soloveychik.
\newblock Central limit theorem for symmetric exchangeable random variables.
\newblock {\em arXiv preprint arXiv:2006.10819}, 2020.

\bibitem{song2017tell}
Jing Song, Satoshi Oyama, and Masahito Kurihara.
\newblock Tell cause from effect: models and evaluation.
\newblock {\em International Journal of Data Science and Analytics}, 4:99--112, 2017.

\bibitem{spirtes2000constructing}
Pater Spirtes, Clark Glymour, Richard Scheines, Stuart Kauffman, Valerio Aimale, and Frank Wimberly.
\newblock Constructing bayesian network models of gene expression networks from microarray data.
\newblock 2000.

\bibitem{wang2023confidence}
Y~Samuel Wang, Mladen Kolar, and Mathias Drton.
\newblock Confidence sets for causal orderings.
\newblock {\em arXiv preprint arXiv:2305.14506}, 2023.

\bibitem{xu2014pooling}
Lele Xu, Tingting Fan, Xia Wu, KeWei Chen, Xiaojuan Guo, Jiacai Zhang, and Li~Yao.
\newblock A pooling-lingam algorithm for effective connectivity analysis of fmri data.
\newblock {\em Frontiers in computational neuroscience}, 8:125, 2014.

\bibitem{zhang2006extensions}
Kun Zhang and Lai-Wan Chan.
\newblock Extensions of ica for causality discovery in the hong kong stock market.
\newblock In {\em International Conference on Neural Information Processing}, pages 400--409. Springer, 2006.

\bibitem{zhang2012identifiability}
Kun Zhang and Aapo Hyvarinen.
\newblock On the identifiability of the post-nonlinear causal model.
\newblock {\em arXiv preprint arXiv:1205.2599}, 2012.

\bibitem{zhang2012inferring}
Xiujun Zhang, Xing-Ming Zhao, Kun He, Le~Lu, Yongwei Cao, Jingdong Liu, Jin-Kao Hao, Zhi-Ping Liu, and Luonan Chen.
\newblock Inferring gene regulatory networks from gene expression data by path consistency algorithm based on conditional mutual information.
\newblock {\em Bioinformatics}, 28(1):98--104, 2012.

\end{thebibliography}

\appendix

\section*{Appendix}

\section{Proof of Theorem 1 in Main Text}

    Recall that $n$ is a fixed subsample size, $s \in \{1,\cdots, S\}$ is the subsample number, $N$ is the dataset sample size, $S_{n,s}$ is the $s^{th}$ subsample of size $n$, $W_{n, s}$ is an indicator for the casual discovery method choosing the hypothesized direction ($X$ causes $Y$) for subsample $S_{n,s}$, and $\hat{p}_n \equiv \frac{1}{S}\sum^S_{s=1}W_{n,s}$ is the estimated rate of detecting $X \rightarrow Y$ for subsample size $n$. We show the CLT and consistency results without loss of generality for $\hat{p}_{n}$ but these results hold any other probability estimate in Algorithm 1 in the main text. 
    
    Let $$Z_{n,S_s} \equiv \frac{W_{n,s} - E[W_{n,s}]}{\sqrt{\sfrac{\Var[W_{n,s}]}{S}}} \in [-1,1]$$

    Note that $Z_{n, S_s}$ is a random variable with only two outcomes. Additionally, as $n$ is fixed, for ease of notation we let $S_i \equiv S_{n, i}$.

\subsection{Preliminary Results and Proofs}
    \setcounter{remark}{1}
    \begin{remark}
        \label{rem:exchange}
        $Z_{n,S_1}, Z_{n,S_2},\cdots,Z_{n,S_s}$ are symmetric exchangeable sequence of random variables.
    \end{remark}
    \begin{proof}
        We want to show $P_{z_{n,S_1}, z_{n,S_2},\cdots,z_{n,S_S}}(z_{n,S_1}, z_{n,S_2},\cdots,z_{n,S_S})$ are symmetric in arguments $z_{n,S_1}, z_{n,S_2},\cdots,z_{n,S_S}$.
        We have that for any $i, j \in \{1,2,\cdots,S\}, P(z_{n,S_i}) = P(z_{n,S_j})$. This is because we are sampling the subsamples $S_{i}$ and $S_{j}$ identically distributed with replacement, so the distribution of $P(z_{n, S_i})$ and $P(z_{n, S_j})$ are the same and the ordering between the subsamples does not affect the joint distribution. So in other words we have that $P_{z_{n,S_1}, z_{n,S_{2}},\cdots,z_{n,S_S}}(z_{n,S_1}, z_{n,S_2},\cdots,z_{n,S_S})$ is symmetric in arguments $z_{n,S_1}, z_{n,S_2},\cdots,z_{n,S_S}$.
    \end{proof}

    \begin{lemma} 
    \label{lem:overlap}
    If $\lim_{N,S \to \infty} S n\log\Big(1-\frac{(nS)(nS - S)}{N(N-nS)^{n-1}}\Big) = 0$ and $N > Sn$, then for any $i, j \in \{1,\cdots,S\}, i\neq j$, for subsamples $S_i$ and $S_j$, $\lim_{S,N \to \infty} P((S_i \cap S_j) = \emptyset) \to 1$. In other words, for any two subsamples, we have that there is asymptotically no overlap between subsamples as $N,S \to \infty$ if $\lim_{N,S \to \infty} S n\log\Big(1-\frac{(nS)(nS - S)}{N(N-nS)^{n-1}}\Big) = 0$ and $N > Sn$.

    \begin{proof}
    Let $\{S_i\}$ denote the number of unique elements in subsample $S_i$ (i.e. separate indices from the dataset of size $N$) and $S_{1,\cdots,S-1} \equiv S_1 \cup S_2 \cup S_3 \cup \cdots \cup S_{S-1}$.

    Then we have that
    $$P((S_i \cap S_j) = \emptyset), \forall i,j \in \{1,\cdots,S\} \Leftrightarrow P((S_1 \cap S_2 \cap S_3 \cap \cdots \cap S_S) = \emptyset)$$
    \begin{align}
        P((S_1 \cap \cdots \cap S_S) = \emptyset) &= P((S_1 \cup \cdots \cup S_{S-1}) \cap S_S = \emptyset \text{ and } S_1 \cap \cdots \cap S_{S-1} = \emptyset) \\
        &= P(S_{1,\cdots,S-1} \cap S_S = \emptyset | S_1 \cap \cdots \cap S_{S-1} = \emptyset)P(S_1 \cap \cdots \cap S_{S-1} = \emptyset) \\
        &= \sum_{k=S-1}^{n(S-1)} \Bigg[P(S_{1,\cdots,S-1}\cap S_S = \emptyset | \{S_{1,\cdots,S-1}\}=k, S_1 \cap \cdots \cap S_{S-1} = \emptyset) \notag\\
        &\times P(\{S_{1,\cdots,S-1}\}=k| S_1 \cap \cdots \cap S_{S-1} = \emptyset)\Bigg] \times P(S_1 \cap \cdots \cap S_{S-1} = \emptyset) \\
        &= \sum_{k=S-1}^{n(S-1)} \Bigg[\Bigg(1-\frac{k}{N}\Bigg)^{n} P(\{S_{1,\cdots,S-1}\}=k| S_1 \cap \cdots \cap S_{S-1} = \emptyset)\Bigg] P(S_1 \cap \cdots \cap S_{S-1} = \emptyset) \\
        &\geq  \Bigg(1-\frac{\sum_{k=S-1}^{n(S-1)}k P(\{S_{1,\cdots,S-1}\}=k| S_1 \cap \cdots \cap S_{S-1} = \emptyset)}{N}\Bigg)^{n} P(S_1 \cap \cdots \cap S_{S-1} = \emptyset)
    \end{align}
    The equality in line 1 can be shown with a diagram studying the unions and intersections of sets. Line 2 follows from the definition of conditional probability. Line 3 follows from the law of total probability. Line 4 follows from $P(S_{1,\cdots,S-1}\cap S_S = \emptyset | \{S_{1,\cdots,S-1}\}=k, S_1 \cap \cdots \cap S_{S-1} = \emptyset) = \left(1-\frac{k}{N}\right)^{n}$; this probability assumes that $N > n(S-1)$ or sufficiently $N > nS$. Line 5 follows from Jensen's inequality: $E[f(k)|X] \geq f(E[k|X]) = f\bigg(\sum_k k p(k|X)\bigg), f(k) = \bigg(1-\frac{k}{N}\bigg)^n$ is convex.

    Then we have that
    \setcounter{equation}{5}
    \begin{align}
        P(\{S_{1,\cdots,S-1}\}=k| S_1 \cap \cdots \cap S_{S-1} = \emptyset) &= \sum_{m=S-2}^{k-1} P(\{S_{1,\cdots,S-2}\}=m \text{ and } \{S_{S-1}\} = k-m | S_1 \cap \cdots \cap S_{S-1} = \emptyset) \\
        &= \sum_{m=S-2}^{k-1}  P(\{S_{S-1}\} = k-m|S_1 \cap \cdots \cap S_{S-1} = \emptyset,\{S_{1,\cdots,S-2}\}=m) \notag \\
        &\times P(\{S_{1,\cdots,S-2}\}=m|S_1 \cap \cdots \cap S_{S-1} = \emptyset) \\
        &= \sum_{m=S-2}^{k-1}{N-m \choose k-m} \bigg(\frac{k-m}{N-m}\bigg)^n P(\{S_{1,\cdots,S-2}\}=m|S_1 \cap \cdots \cap S_{S-1} = \emptyset)\\
        &> {N-k+1 \choose 1} \bigg(\frac{1}{N-k+1}\bigg)^n  \sum_{m=S-2}^{k-1} P(\{S_{1,\cdots,S-2}\}=m|S_1 \cap \cdots \cap S_{S-1} = \emptyset) \\
        &= \frac{1}{(N-k+1)^{n-1}} 
    \end{align}
    Line 7 follows from the definition of conditional probability. Line 8 follows from $P(\{S_{S-1}\} = k-m|S_1 \cap \cdots \cap S_{S-1} = \emptyset,\{S_{1,\cdots,S-2}\}=m) = {N-m \choose k-m}\left(\frac{k-m}{N-m}\right)^n$ since we first choose the $k-m$ elements from pool of $N-m$, then the probability of ``success" is $\left(\frac{k-m}{N-m}\right)$ for $n$ draws. Line 9 follows from ${N-m \choose k-m}\left(\frac{k-m}{N-m}\right)^n$ being the smallest for the largest value of $m$, which is $k-1$. Line 10 follows from $\sum_{m=S-2}^{k-1} P(\{S_{1,\cdots,S-2}\}=m|S_1 \cap \cdots \cap S_{S-1} = \emptyset) = 1$. 
    
    Now we can use line 10 to plug into line 5,
    \begin{align*}
        \Bigg(1-\frac{\sum_{k=S-1}^{n(S-1)}k P(\{S_{1,\cdots,S-1}\}=k| S_1 \cap \cdots \cap S_{S-1} = \emptyset)}{N}\Bigg)^{n} &> \Bigg(1-\frac{\sum_{k=S-1}^{n(S-1)}\frac{k}{(N-k+1)^{n-1}}}{N}\Bigg)^{n} \\
        \implies P(S_{1,\cdots,S-1} \cap S_S = \emptyset | S_1 \cap \cdots \cap S_{S-1} = \emptyset)P(S_1 \cap \cdots \cap S_{S-1} = \emptyset) &> \Bigg(1-\frac{\sum_{k=S-1}^{n(S-1)}\frac{k}{(N-k+1)^{n-1}}}{N}\Bigg)^{n} \\
        &\times P(S_1 \cap \cdots \cap S_{S-1} = \emptyset) 
    \end{align*}
    Similarly for $S-1$, we have that 
    $$P(S_1 \cap \cdots \cap S_{S-1}  = \emptyset) > \Bigg(1-\frac{\sum_{k=S-2}^{n(S-2)}\frac{k}{(N-k+1)^{n-1}}}{N}\Bigg)^{n} P(S_1 \cap \cdots \cap S_{S-2} = \emptyset)$$
    We can bound $P(S_1 \cap \cdots \cap S_{S-2}  = \emptyset)$ similarly and this would recursively hold until $S=2$.
    
    So we have,
    \setcounter{equation}{10}
    \begin{align}
        P((S_1 \cap \cdots \cap S_S) = \emptyset) &> \Bigg(1-\frac{\sum_{k=1}^{n}\frac{k}{(N-k+1)^{n-1}}}{N}\Bigg)^{n} \times \cdots \times \Bigg(1-\frac{\sum_{k=S-1}^{n(S-1)}\frac{k}{(N-k+1)^{n-1}}}{N}\Bigg)^{n}\\
        &= \prod_{i=1}^{S-1}\Bigg(1-\frac{\sum_{k=i}^{n \times i}\frac{k}{(N-k+1)^{n-1}}}{N}\Bigg)^{n} \\
        &> \prod_{i=1}^{S-1}\Bigg(1-\frac{\frac{(n\times i)(n \times i - i + 1)}{(N-n\times i+1)^{n-1}}}{N}\Bigg)^{n} \\
        &= \prod_{i=1}^{S-1}\Bigg(1-\frac{(n\times i)(n \times i - i + 1)}{N(N-n\times i+1)^{n-1}}\Bigg)^{n} \\
        &> \Bigg(1-\frac{(n(S-1))(n(S-1) - (S-1) + 1)}{N(N-n(S-1)+1)^{n-1}}\Bigg)^{n(S-1)} 
    \end{align}
    Line 13 follows from the sum being the largest for $k = n \times i$, so the fraction is largest and closest 1 when we have the largest value throughout the sum. Line 15 follows since the fraction is closest to 1 (largest) for $i = (S-1)$.
    
    Now if we look as $N,S \to \infty$ we have that
    \setcounter{equation}{15}
    \begin{align}
        \lim_{N,S \to \infty} P((S_i \cap S_j) = \emptyset) &= \lim_{N,S \to \infty} P((S_1 \cap \cdots \cap S_S) = \emptyset)\\
        &= \lim_{N,S \to \infty} e^{(S-1)n\log\Big(1-\frac{(n(S-1))(n(S-1) - (S-1) + 1)}{N(N-n(S-1)+1)^{n-1}}\Big)} \\
        &= \lim_{N,S \to \infty} e^{S n\log\Big(1-\frac{(nS)(nS - S)}{N(N-nS)^{n-1}}\Big)} \\
        &= e^{\lim_{N,S \to \infty} S n\log\Big(1-\frac{(nS)(nS - S)}{N(N-nS)^{n-1}}\Big)}\\ 
        &= e^0 = 1
    \end{align}
    Line 20 follows since $\lim_{N,S \to \infty} S n\log\Big(1-\frac{(nS)(nS - S)}{N(N-nS)^{n-1}}\Big) = 0$ and $N > Sn$.
    \end{proof}
\end{lemma}

\subsection{Main Text Theorem 1 Proof}

\begin{enumerate}
    \item \textbf{Proof of consistency:}

\setcounter{equation}{0}
\begin{proof}
    Let $W_{n,s,1}$ denote the part of $W_{n,s}$ that is free of any overlap between subsamples and $W_{n,s,2}$ denote the part of $W_{n,s}$ that contains overlap between subsamples. Then we have
    \begin{align}
    \lim_{N,S \to \infty} P\left(\left|\hat{p}_{n} - p_n\right| < \epsilon \right) &= \lim_{N,S \to \infty} P\left(\left|\frac{1}{S}\sum^S_{s=1}W_{n,s} - E[W_{n,s}]\right| < \epsilon \right) \\
    &= 1-\lim_{N,S \to \infty} P\left(\left|\frac{1}{S}\sum^S_{s=1}W_{n,s} - E[W_{n,s}]\right| \geq \epsilon \right) \\
    &\geq 1- \lim_{N,S \to \infty}\frac{\Var\left(\frac{1}{S}\sum^S_{s=1}W_{n,s}\right)}{\epsilon^2} \\
    &= 1- \lim_{N,S \to \infty}\frac{\Var\left(\frac{1}{S}\sum^S_{s=1} W_{n,s,1} + \frac{1}{S}\sum^S_{s=1} W_{n,s,2} \right) }{\epsilon^2} \\
    &= 1- \lim_{N,S \to \infty}\frac{\Var\left(\frac{1}{S}\sum^S_{s=1} W_{n,s,1}\right) + \Var\left(\frac{1}{S}\sum^S_{s=1} W_{n,s,2} \right) }{\epsilon^2} \\
    &\geq 1- \lim_{N,S \to \infty}\frac{\Var\left(\frac{1}{S}\sum^S_{s=1} W_{n,s,1}\right) + \Var\left(W_{n,1,2} \right) }{\epsilon^2} \\
    &= 1- \lim_{N,S \to \infty}\frac{\Var\left(\frac{1}{S}\sum^S_{s=1} W_{n,s,1} \right) }{\epsilon^2} \\
    &= 1- \lim_{N,S \to \infty}\frac{\Var\left(W_{n,1,1} \right) }{S\epsilon^2} \\
    &= 1
    \end{align}

    Line 3 follows from Chebyshev's. Line 5 follows from the independence of $\Var\left(\frac{1}{S}\sum^S_{s=1} W_{n,s,1}\right)$ and $\Var\left(\frac{1}{S}\sum^S_{s=1} W_{n,s,2} \right)$ by the construction of $W_{n,s,1}$ and $W_{n,s,2}$. Line 6 follows from $\Var\left(\frac{1}{S}\sum^S_{s=1} W_{n,s,2} \right) \leq \Var(W_{n,1,2})$ since $W_{n,s,2}$ are identically distributed. Line 7 follows from Lemma \ref{lem:overlap}; we have that if $\lim_{N,S \to \infty} S n\log\Big(1-\frac{(nS)(nS - S)}{N(N-nS)^{n-1}}\Big) = 0$ and $N > Sn$, then we have asymptotically no overlap between subsamples as $N,S \to \infty$, which implies $\lim_{N,S \to \infty} \Var(W_{n,1,2}) = 0$. Line 9 follows since $\Var(W_{n,1,1})$ is finite, so $\lim_{N,S \to \infty} \frac{\Var(W_{n,1,1})}{S\epsilon^2} = 0$.
    \end{proof}

    \item \textbf{Proof of CLT result}:
    
    This CLT result comes directly from Soloveychik's result in Theorem 1 \cite{soloveychik2020central}. 

    We have, as shown in Remark \ref{rem:exchange}, that $Z_{n,S_s}$ for different $s$ are symmetric exchangeable random variables.

    Let $\{k_S\}_{S=1}^\infty$ and $\{m_S\}_{S=1}^\infty$ be two sequences of natural numbers such that $k_S < m_S$ and $$\frac{k_S}{m_S} \rightarrow \gamma \in [0,1)$$

    We are dealing with finite exchangeable sequences so we consider triangular arrays of symmetric row-wise exchangeable random variables of the form $\{Z_{n,S_s}\}_{S,s=1}^{\infty, m_S}$. We have that the sequences are symmetric and row-wise exchangeable because for each row, the ordering of the random variables doesn't effect total joint distribution and the switching of rows also doesn't effect the total joint distribution.

    Let $N \to \infty$ and assume for fixed $n$ that $\lim_{N,S \to \infty} S n\log\Big(1-\frac{(nS)(nS - S)}{N(N-nS)^{n-1}}\Big) = 0$ and $N > Sn$. Now for the Soloveychik's CLT result \cite{soloveychik2020central} to hold we need the following conditions:
    \begin{enumerate}
    \item $E[Z_{n,S_1} Z_{n,S_2}] \to 0$ as $S\to \infty$
    \item $\max_{1 \leq s \leq m_S} \frac{|Z_{n,S_s}|}{\sqrt{m_S}} \overset{p}{\to} 0$ (as $S \rightarrow \infty)$
    \item $\frac{1}{m_S}\sum_{s=1}^{m_S} Z_{n,S_s}^2 \overset{p}{\to} 1$, (as $S \rightarrow \infty$)
\end{enumerate}
\textbf{\textit{Proof Condition (a) holds:}}
\begin{proof}
Let $S_{11}$ denote the part of subsample $S_1$ that is disjoint of $S_2$. Let $S_{22}$ denote the part of subsample $S_2$ that is disjoint of $S_1$. Let $S_{12}$ denote the overlap between subsamples $S_1$ and $S_2$. Then we have

\setcounter{equation}{0}
\begin{align}
\lim_{N,S \to \infty} E[Z_{n,S_1} Z_{n,S_2}] &= \lim_{N,S \to \infty}E[(Z_{n,S_{11}}+Z_{n,S_{12}})(Z_{n,S_{22}}+Z_{n,S_{12}})] \\
&= \lim_{N,S \to \infty}\left(E[Z_{n,S_{11}}Z_{n,S_{22}}]+E[Z_{n,S_{11}}Z_{n,S_{12}}] + E[Z_{n,S_{12}}Z_{n,S_{22}}] + E[Z_{n,S_{12}}^2] \right)\\
&= \lim_{N,S \to \infty}E[Z_{n,S_{11}} Z_{n,S_{22}}] \\
&= \lim_{N,S \to \infty}E[Z_{n,S_{11}}]E[Z_{n,S_{22}}] \\
&= 0
\end{align}

Line 3 follows from Lemma \ref{lem:overlap}, meaning that there is asymptotically no overlap between $S_1$ and $S_2$; in other words, if $\lim_{N,S \to \infty} S n\log\Big(1-\frac{(nS)(nS - S)}{N(N-nS)^{n-1}}\Big) = 0$ and $N > Sn \implies \lim_{S,N \to \infty} E[Z_{n,S_{11}}Z_{n,S_{12}}] = E[Z_{n,S_{12}}Z_{n,S_{22}}] = E[Z_{n,S_{12}}^2] = 0$. Line 4 follows since we have two disjoint subsamples, we have that $Z_{n,S_{11}}$ and $Z_{n,S_{22}}$ are independent of each other asymptotically. Line 5 follows since $E[Z_{n,S_{11}}] =0, E[Z_{n,S_{22}}] =0$.

\end{proof}

\setcounter{equation}{0}
\textit{\textbf{Proof Condition (b) holds:}}
\begin{proof}
\begin{align}
    \max_{1 \leq s \leq m_S} \frac{|Z_{n,S_s}|}{\sqrt{m_S}} &= \lim_{S \to \infty} \max_{1 \leq s \leq m_S} P\bigg(\frac{|Z_{n,S_s}|}{\sqrt{m_S}}  \geq \epsilon\bigg), \forall \epsilon > 0 \\
    &= \lim_{S \to \infty} \max_{1 \leq s \leq m_S} P(|Z_{n,S_s}|  \geq \epsilon \sqrt{m_S}), \forall \epsilon > 0 \\
    &\leq \lim_{S \to \infty} \max_{1 \leq s \leq m_S} \frac{\Var(Z_{n,S_s})}{\epsilon^2 m_S}, \forall \epsilon > 0 \\
    &= 0
\end{align}

Line 3 holds by Chebyshev's and $E[Z_{n,S_s}] = 0$. Line 4 holds because we know that $\Var(Z_{n,S_s})$ is finite and since $S \to \infty, m_S \to \infty$ since it is the infinite index to a natural number sequence which would be $\infty$. Then we have that the limit is 0.
\end{proof}

\textit{\textbf{Proof Condition (c) holds:}}
\begin{proof}
We want to show $$\frac{1}{m_S}\sum_{s=1}^{m_S} Z_{n,S_s}^2 \overset{p}{\to} 1, S \rightarrow \infty, N \to \infty$$

We have by Lemma \ref{lem:overlap} that for fixed $n$ if $\lim_{N,S \to \infty} S n\log\Big(1-\frac{(nS)(nS - S)}{N(N-nS)^{n-1}}\Big) = 0$ and $N > Sn$, then the probability of no overlap between the subsamples of the i.i.d data would go to 1 asymptotically. This would mean that if we have that $\lim_{N,S \to \infty} S n\log\Big(1-\frac{(nS)(nS - S)}{N(N-nS)^{n-1}}\Big) = 0$ and $N > Sn$, then we would have asymptotically i.i.d samples. 

Then by WLLN, if we have that $S \to \infty$ (which implies $m_S \to \infty$), we have that

\begin{align*}
    \frac{1}{m_S}\sum_{s=1}^{m_S} Z_{n,S_s}^2 &\overset{p}{\to} E[Z_{n,S_s}^2] \\
    &= \Var(Z_{n,S_s}) + (E[Z_{n,S_s}])^2 \\
    &= 1 + 0 \\
    &= 1
\end{align*}
\end{proof}

Since we have that all three conditions hold, we have that the CLT results hold for our example as $N \to \infty, S \to \infty$ as long as $\lim_{N,S \to \infty} S n\log\Big(1-\frac{(nS)(nS - S)}{N(N-nS)^{n-1}}\Big) = 0$ and $N > Sn$ for fixed $n$.
\end{enumerate}

\section{Validating Asymptotic Normality of CDDR Diagnostic (Remark 1 from Main Text)}

Without loss of generality, we will use simulations to check the validity of asymptotic normality of the estimated rate of detecting the hypothesized direction for the test-based approach using the Sen and Sen test \cite{sen_testing_2014}. We denote the true direction as $X \rightarrow Y$. Then the hypothesized direction would align with the true direction. Let $W_{n_j,s}$ be an indicator function indicating whether the test-based approach rejects $H_Y^0$ and fails to reject $H_X^0$ in Equation (6) in the main text for subsample $S_{n_j,s}$. Then we can estimate the rate of detecting $X \rightarrow Y$ for subsample size $n_j$ as the following:
$$\hat{p}_{n_j}\equiv \frac{1}{S}\sum^S_{s=1}W_{n_j,s}$$

We will validate the asymptotic normality by showing we can approximate the true variability of $\hat{p}_{n_j}$ using the normal approximation. We estimate the true variability of $\hat{p}_{n_j}$ by replicating its estimation procedure (shown in Algorithm 1 in the main text) $M$ times. In our simulations, we let $M = 100$ and our level of significance, $\alpha$, be $0.05$. We conduct simulations for two settings where our number of subsamples, $S$, is $100$ and $500$ respectively. This corresponds to cases where our sufficient conditions are satisfied and where they are not. We let our total sample size be $N = 10000$ and our subsample sizes ($n_j$'s) range from 20 to 120 subsamples. We simulate $X$ from a truncated exponential distribution and we let $Y = sign(X-a)|X-a|^{1.25}*\beta + \epsilon$, such that it is slightly non-linear. We sample $\epsilon$ from a Gaussian Mixture Model (GMM) with 3 mixtures. For our simulations, we estimate the standard error and confidence intervals using the following formulas:

\begin{align*}
    \hat{SE}_{n_j} &= \sqrt{\frac{\hat{p}_{n_j}(1-\hat{p}_{n_j})}{S}} \\
    \text{CI}_{\text{lower}, n_j} &= \hat{p}_{n_j}-z_{\alpha}*\hat{SE}_{n_j} \\
    \text{CI}_{\text{upper},n_j} &= \hat{p}_{n_j}+z_{\alpha}*\hat{SE}_{n_j} \\
\end{align*}

Note that $z_{\alpha}$ is the z-score corresponding to a significance level $\alpha$. For each setting ($S = 100, 500$), we will estimate the bias between these quantities and the standard deviation (denoted as $SD_{n_j}$), $97.5^{th}\%$ quantile (denoted as $Q_{\text{upper},n_j}$), and $2.5^{th}\%$ quantile (denoted as $Q_{\text{lower},n_j}$) of the estimated sampling distribution of $\hat{p}_{n_j}$ respectively:

\begin{align*}
    &\frac{1}{M}\sum_{m=1}^M (\hat{SE}_{n_j,m}-SD_{n_j}) \\
    &\frac{1}{M}\sum_{m=1}^M (\text{CI}_{\text{lower},  n_j, m}-Q_{\text{lower},n_j}) \\
    &\frac{1}{M}\sum_{m=1}^M (\text{CI}_{\text{upper},  n_j, m}-Q_{\text{upper},n_j})
\end{align*}

\begin{figure}
    \centering
    \includegraphics[width=175mm]{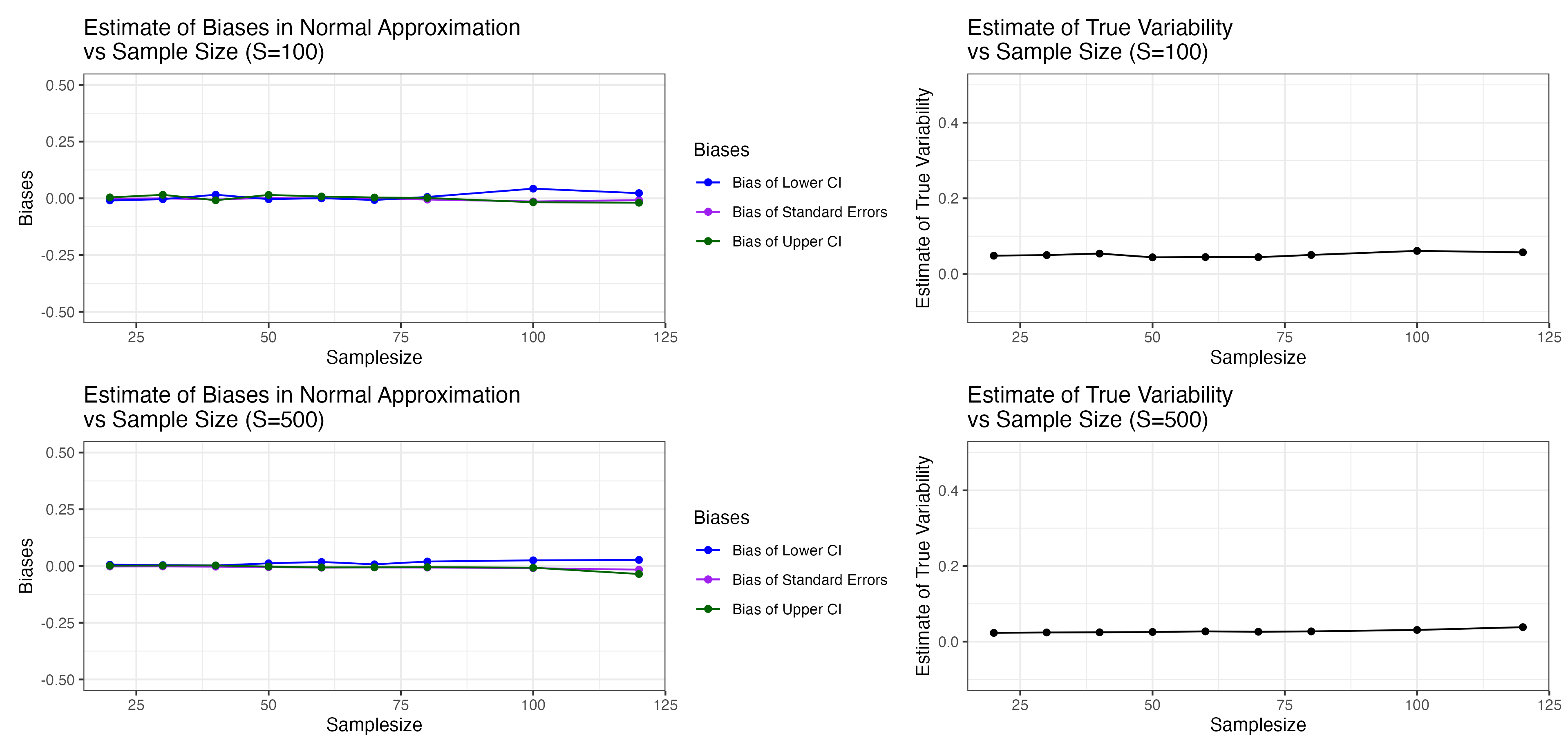}
    \caption{These plots demonstrate that we can use the normal approximation for confidence intervals both in cases where the sufficient conditions in Theorem 1 in the main text both hold and do not hold. First column plots the biases between the sample and estimated true quantities. The second column plots the variability of the estimated sampling distribution of $\hat{p}_{n_j}$. The first row is for the first setting (S=100), where the sufficient conditions hold, and the second row is for the second setting (S=500), where the sufficient conditions do not hold. We see that for both settings, the biases are very small and the variability is small and constant across subsample sizes ($n_j$). }
    \label{fig:sim_clt}
\end{figure}

We plot these estimated biases as well as the variability of the estimated sampling distribution of $\hat{p}_{n_j}$ for both settings in Figure \ref{fig:sim_clt}. Under both settings, we see that the biases are small, even though in Setting 2, the sufficient conditions do not hold. This provides evidence that the normal approximation provides good estimates of the confidence intervals in practice both in cases where the sufficient condition holds and does not hold. We also have that since $\hat{p}_{n_j}$ is a proportion, the variability does not depend on $n_j$, thus indicating why the variability of the estimated sampling distribution is constant across different values of $n_j$. This is true regardless of whether our sufficient condition holds or not. Furthermore, we note that there is little variability in the estimated sampling distribution of $\hat{p}_{n_j}$. The variability of $\hat{p}_{n_j}$ would be maximized if there was an equal probability of choosing either direction ($P(X \rightarrow Y) = P(Y \rightarrow X) = 0.5$). Then, we would have $SD(p_{n_j}) = 0.5$. When comparing our estimates to this value, we can see that they are small in comparison. This indicates that our CDDR diagnostic will be relatively smooth as seen in our simulations and real data analysis (Sections 3 and 4 in the main text). 

\section{Smaller Sample Size Simulations}

\begin{figure}
    \centering
    \includegraphics[width=160mm]{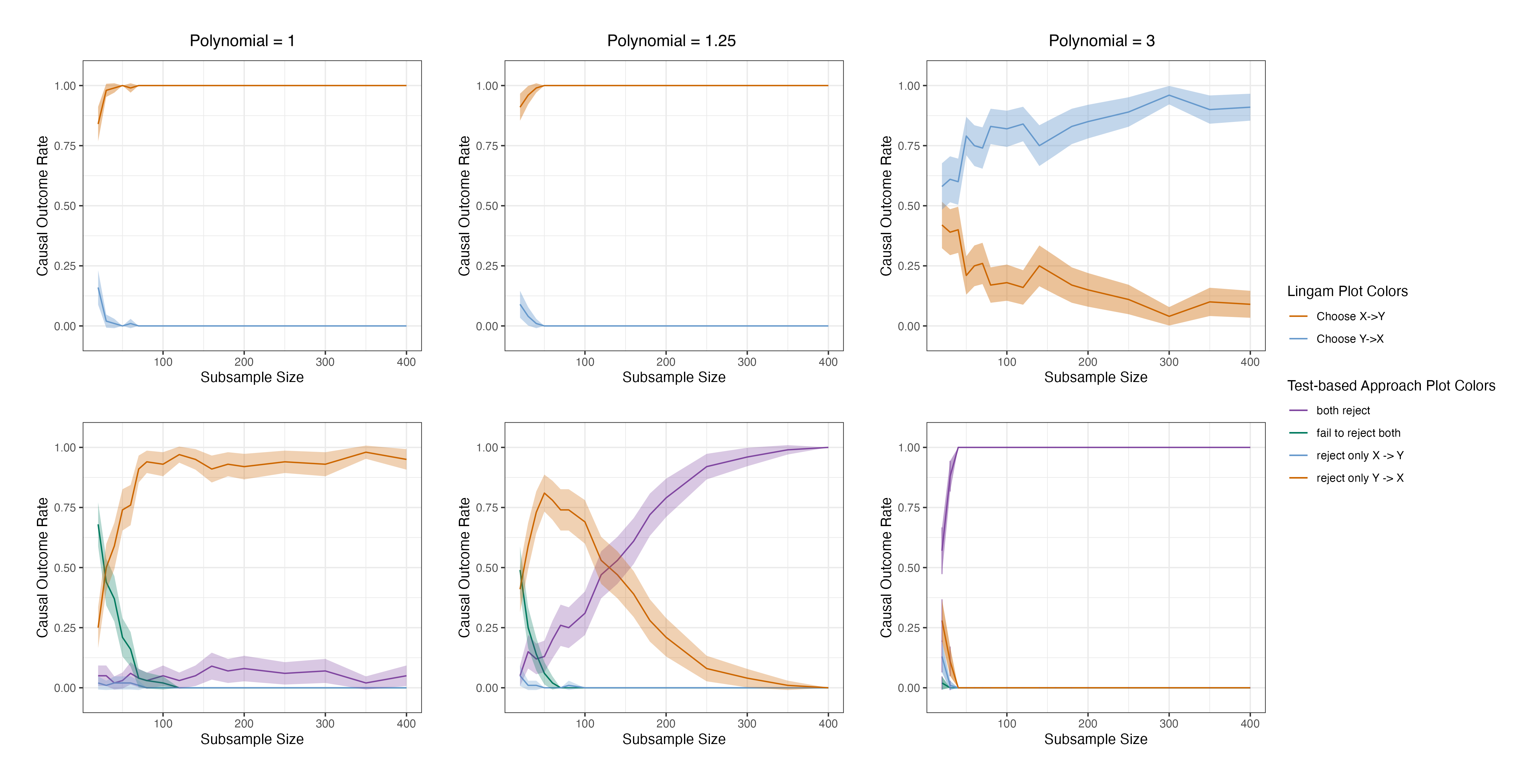}
    \caption{Linearity Assumption Violated for n=400. In each plot, the x-axis represents the subsample sizes, which are less than or equal to $N$, and the y-axis represents the causal outcome rates. The first row represents running the LiNGAM algorithm and the second row represents running the test-based approach. The first column is when Polynomial = 1, second column is when Polynomial = 1.25, and third column is when Polynomial = 3.}
    \label{fig:sim_lins_small}
\end{figure}

Figure \ref{fig:sim_lins_small} shows the results from our simulations assessing the linearity assumption for the smaller sample size (n = 400). These results are very similar to the larger sample size case.

\begin{figure}
    \centering
    \includegraphics[width=160mm]{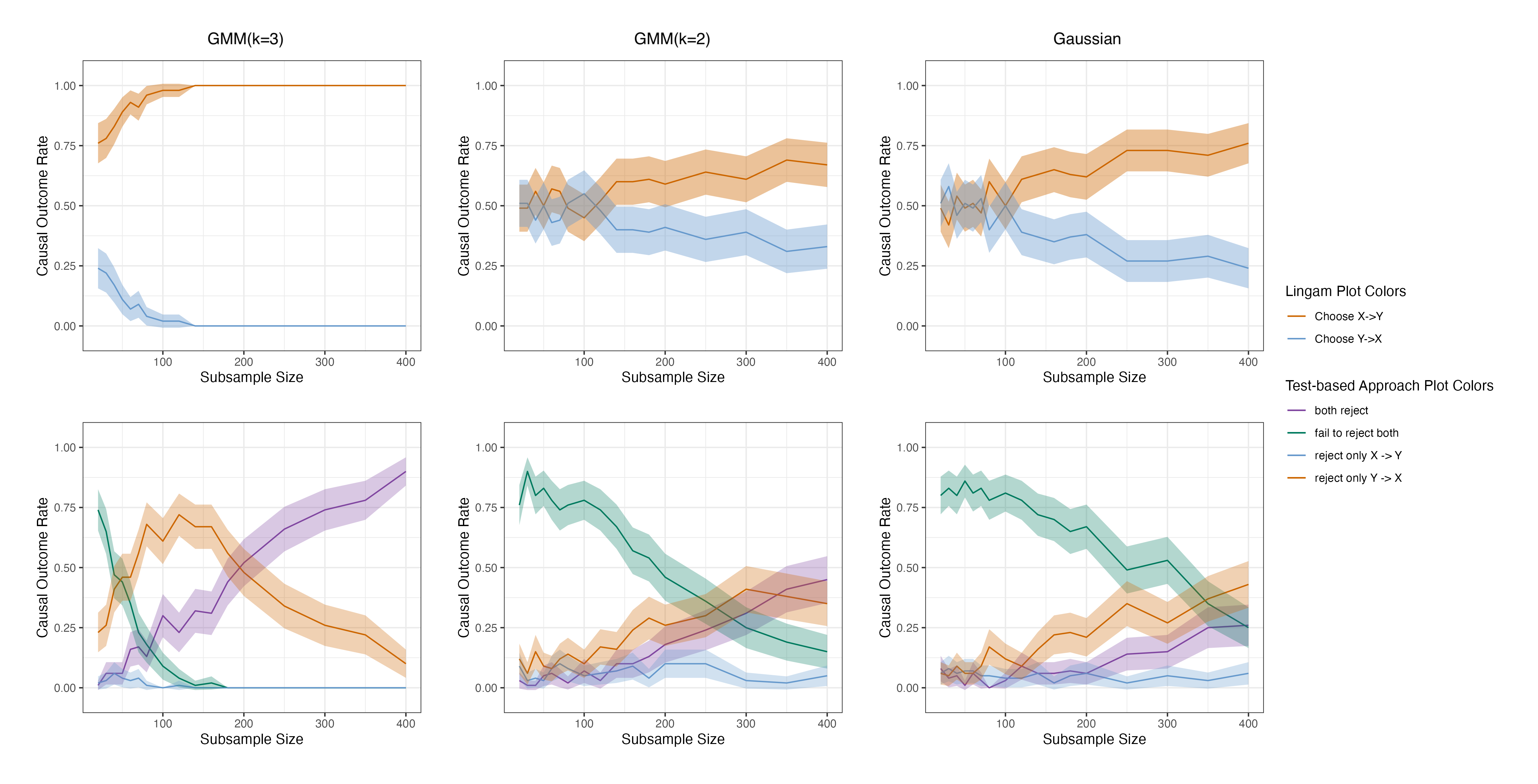}
    \caption{Non-Gaussianity Assumption Violated for n=400. The first row represents running the LiNGAM algorithm and the second row represents running the test-based approach. The first column is when use Guassian errors and X, second column is when use GMM(k=1) for the errors and Normally distributed X, and third column is when use GMM(k=3) for the errors and Normally distributed X.}
    \label{fig:sim_gauss_small}
\end{figure}

Figure \ref{fig:sim_gauss_small} shows the results from our simulations assessing the non-Gaussianity assumption for the smaller sample size (n = 400). In each plot, the x-axis represents the subsample sizes, which are less than or equal to $N$, and the y-axis represents the causal outcome rates. These results are very similar to the larger sample size case, however, because of the small sample size as well as the non-Gaussianity assumption violations, we fail to reject for most but not all subsample sizes for the severe assumption violation case.

\section{LiNGAM and the Test-based Approach under Scientifically Derived Transformations}

We will first show that we can identify LiNGAM and the test-based approach when we have a pre-specified transformation determined by science: $f(X)$.

For both methods, we compare the equations
\setcounter{equation}{0}
\begin{equation}
    \label{eq:lingam_ex_xy}
    Y = \beta f(X) + \epsilon
\end{equation}
and 
\begin{equation}
    \label{eq:lingam_ex_yx}
    f(X) = \gamma Y + \eta,
\end{equation}
 which we will show is a trivial extension of LiNGAM with the non-Guassianity assumption on $f(X)$ rather than $X$.

Both methods still require the following assumptions for identification:

\begin{enumerate}
    \item There exists a linear relationship between $f(X)$ and $Y$,
    \item At least one of the variables $f(X),Y$, or the error term is non-Gaussian,
    \item $f(X)$ and $Y$ have an acyclic relationship,
    \item $f(X)$ and $Y$ are unconfounded, meaning they have no common causes
    \item All involved random variables are i.i.d
\end{enumerate}

However, as seen, we have a relaxation on the linearity and a modified non-Gaussianity assumption. Similar to LiNGAM we have the following two propositions:

\begin{proposition} 
\label{prop:lingam_extend}
Let $X$ and $Y$ be two random variables, and $f(X)$ be a pre-specified transformation determined by science. Assume that 
\begin{equation}
    Y = \beta f(X) + \epsilon, \quad \epsilon \perp f(X), \beta \neq 0
\end{equation}
holds. Then we can reverse the process, i.e. there exists $\gamma \in \mathbb{R}$ and a noise $\eta$ such that
\begin{equation}
    f(X) = \gamma Y + \eta, \quad \eta \perp Y
\end{equation}
if and only if $f(X),Y,\epsilon,\eta$ are Gaussian distributed.
\end{proposition}

\begin{proof}
This follows directly from the LiNGAM identification results. We can prove this using Theorem 2.10 from Peters~\cite{peters_2008} where we plug in $f(X)$ for X.
\end{proof}

\begin{proposition}
\label{prop:lingam_extend_id}
Let $X$ and $Y$ be two random variables, and $f(X)$ be a pre-specified transformation determined by science. Assume that (\ref{eq:lingam_ex_xy}) or
$$Y = \beta f(X) + \epsilon, \quad \epsilon \perp f(X), \beta \neq 0$$
holds. Then if we have that the error terms are non-Gaussian and/or that $f(X),Y$ are non-Gaussian, then we have that (\ref{eq:lingam_ex_xy}) and (\ref{eq:lingam_ex_yx}) are distinguishable.
\end{proposition}

\begin{proof}

This follows directly from the LiNGAM identification results. We can prove this using Proposition 1 from Shimizu et al.~\cite{shimizu_use_2008} where we plug in $f(X)$ for $X$.

\end{proof}

Thus we can identify the direction under this pre-specified transformation for both LiNGAM and the test-based approach.

\end{document}